\documentclass[12pt,a4paper]{article}
\pdfoutput=1
\usepackage{jheppub}

\usepackage{enumerate}
\usepackage{graphicx}
\usepackage{caption} 
\usepackage{subcaption} 

\usepackage[dvipsnames]{xcolor}

\usepackage{bm}
\usepackage{amssymb}
\usepackage{amsmath}
\usepackage{cancel}
\usepackage{epigraph}

\newcommand{\captionfonts}{\small}

\makeatletter  
\long\def\@makecaption#1#2{%
  \vskip\abovecaptionskip
  \sbox\@tempboxa{{\captionfonts #1: #2}}%
 \ifdim \wd\@tempboxa >\hsize
    {\captionfonts #1: #2\par}
  \else
    \hbox to\hsize{\hfil\box\@tempboxa\hfil}%
  \fi
  \vskip\belowcaptionskip}
\makeatother   

\usepackage{fancyhdr}
\usepackage{datetime}

\usepackage{mciteplus}

\PassOptionsToPackage{ 
     colorlinks=true,
     linkcolor=myBlue,
     urlcolor=blue,
     filecolor=black,
     citecolor=myRed,
}{hyperref}
      
\usepackage[all]{hypcap}     

\makeatletter
\renewcommand\section{\@startsection {section}{1}{\z@}%
                                   {-3.5ex \@plus -1ex \@minus -.2ex}
                                   {2.3ex \@plus.2ex}%
                                   {\normalfont\Large\bfseries}}
\renewcommand\subsection{\@startsection{subsection}{2}{\z@}%
                                     {-3.25ex\@plus -1ex \@minus -.2ex}%
                                     {1.5ex \@plus .2ex}%
                                     {\normalfont\bfseries}}
\renewcommand\subsubsection{\@startsection{subsubsection}{3}{\z@}%
                                     {-2.5ex\@plus -1ex \@minus -.2ex}%
                                     {1.25ex \@plus .2ex}%
                                     {\normalfont\textit}}
\makeatother

\def\sl{\text{sl}}
\def \su{\text{su}}

\newcommand{\n}{{\mathfrak n}}
\newcommand{\m}{{\mathfrak m}}

\newcommand{\msl}{m_{\sl}}
\newcommand{\msu}{m_{\su}}
\newcommand{\wsl}{w_{\sl}}
\newcommand{\wsu}{w_{\su}}

\newcommand{\bmsl}{\bar{m}_{\sl}}
\newcommand{\bmsu}{\bar{m}_{\su}}
\newcommand{\bwsl}{\bar{w}_{\sl}}
\newcommand{\bwsu}{\bar{w}_{\su}}
\newcommand{\jsl}{j_{\sl}}
\newcommand{\jsu}{j_{\su}}

\def\Msu{M_{\su}}
\def\bMsu{\bar M_{\su}}
\def\Jsu{{\mathsf J}_{\su}}
\def\Jsl{{\mathsf J}_{\sl}}
\def\Ntot{{\mathsf N}_{\rm tot}}
\def\ntot{{\mathsf n}_{\rm tot}}

\newcommand{\sxi}{\mathcal{S}}
\newcommand{\dzeta}{\mathcal{D}}

\newcommand{\eq}[1]{\eqref{#1}}

\def\ads#1{\ensuremath{\text{AdS}_{#1}\!}}

\newcommand{\ket}[1]{|#1 \rangle}

\def\mathbi#1{\textbf{\em #1}}
\def\tight#1{\! #1 \!}

\def\({\left(}
\def\){\right)}
\def\[{\left[}
\def\]{\right]}

\def\sltwo{{SL(2,\IR)}}
\def\sutwo{{SU(2)}}
\def\uone{U(1)}

\def\mbar{\bar m}

\def\ie{{i.e.}}
\def\eg{{e.g.}}

\def\eff{{\rm eff}}

\def\gstr{g_{\textit s}}

\def\nfive{{n_5}}
\def\nfivetil{{\tilde n_5}}
\def\nfivehat{{\hat n_5}}
\def\none{{n_1}}

\def\NL{\hat N_L}
\def\NR{\hat N_R}

\def\k{\ensuremath{\mathsf{k}}}

\def\sst#1{\scriptscriptstyle{#1}}

\def\X{{\mathsf X}}

\def\x1x2{$x^1$-$x^2$}

\def\ydual{{\hat y}}
\def\Ry{R}

\def\mub{{\boldsymbol\mu}}

\def\jdual{ \jmath^{\sst\vee} }
\def\Phihat{\widehat\Phi}
\def\Psihat{\widehat\Psi}
\def\zhat{\hat z}

\def\what{{\hat w}}

\def\ntil{{\mathsf n}}
\def\bntil{{\bar{\mathsf n}}}
\def\nytil{{\mathsf n}_y}

\def\sst{\scriptscriptstyle}
\def\half{\frac12}
\def\coeff#1#2{{\textstyle \frac{#1}{#2}}}
\def\hf{\coeff12}

\def\One{{\hbox{1\kern-1mm l}}}

\def\Dbar{{\bar D}}

\def\barray{\begin{array}}
\def\earray{\end{array}}
\def\be{\begin{equation}}
\def\ee{\end{equation}}
\def\bea{\begin{eqnarray}}
\def\eea{\end{eqnarray}}
\def\bal{\begin{align}}
\def\eal{\end{align}}
\def\nn{\nonumber}

\newcommand{\bR}{{\mathbb R}}
\newcommand{\bS}{{\mathbb S}}
\newcommand{\bT}{{\mathbb T}}
\newcommand{\bZ}{{\mathbb Z}}

\def\IN{\mathbb{N}}

\def\IR{\mathbb{R}}

\def\IZ{\mathbb{Z}}
\def\cA{{\cal A}}

\def\cC{{\cal C}}
\def\cD{{\cal D}}

\def\cG{{\cal G}}

\def\cM{{\cal M}}
\def\cN{{\cal N}}
\def\cO{{\cal O}}

\def\cR{{\cal R}}
\def\cS{{\cal S}}

\definecolor{cardinal}{rgb}{0.6,0,0}
\definecolor{darkgreen}{rgb}{0,0.4,0}
\definecolor{golden}{rgb}{0.92, 0.7, 0}
\definecolor{midnight}{rgb}{0, 0, 0.5}
\definecolor{darkblue}{rgb}{0, 0, 0.7}

\numberwithin{equation}{section}

\mathchardef\mhyphen="2D

\def\cA{{\cal A}}  \def\cC{{\cal C}}
\def\cD{{\cal D}}  
\def\cG{{\cal G}} \def\cH{{\cal H}} 
  
\def\cM{{\cal M}} \def\cN{{\cal N}} \def\cO{{\cal O}}
  \def\cR{{\cal R}}
\def\cS{{\cal S}}  
  
\def\cY{{\cal Y}}

\def\stJ{{\mathbi J}}
\def\stL{{\mathbi L}}

\def\one{{\hbox{\kern+.5mm 1\kern-.8mm l}}}
\def\zero{{\hbox{0\kern-1.5mm 0}}}

\def\d{ \partial }

\def\tha{\tfrac{1}{2}}

\begin{document}

\title{String dynamics in NS5-F1-P geometries} 

\author{Emil J. Martinec$^a$, Stefano Massai$^a$ {\it and}\,
  David Turton$^b$}

\vspace{0.85 cm}

\affiliation[a]{
Enrico Fermi Institute and Dept. of Physics \\
5640 S. Ellis Ave.,
Chicago, IL 60637-1433, USA 
}

\affiliation[b]{
Mathematical Sciences and STAG Research Centre, University of Southampton, \\
Highfield, Southampton, SO17 1BJ, UK
}

 \emailAdd{ejmartin@uchicago.edu}
 \emailAdd{massai@uchicago.edu}
 \emailAdd{d.j.turton@soton.ac.uk}

\vspace{0.7cm} 

\abstract{
String theory dynamics on certain fivebrane supertube backgrounds is described by an exactly solvable null-gauged WZW model.  We use this description to compute the spectrum of closed string excitations on the three-charge non-supersymmetric solution found by Jejjala, Madden, Ross and Titchener, as well as its supersymmetric limit. The low-lying spectrum matches that of supergravity modes in the effective geometry and exhibits an underlying group-theoretic structure.  Winding sectors describe strings carrying the same charges as the background; processes whereby strings turn into flux or vice-versa are mediated by large gauge transformations on the worldsheet.  The S-matrix of such wound strings probes microstructure of the fivebrane source that is hidden in the supergravity approximation.
}
%

\maketitle

\baselineskip=15pt
\parskip=3pt


\setcounter{footnote}{0}


\section{Introduction and Discussion}

String theory exhibits a rich phase structure, encompassing the wealth of states of gauge theory interacting with matter, and much more.  One intriguing aspect of this structure is the collection of topological transitions in which matter sources become fluxes threading topological cycles.  Prime examples of this phenomenon are the AdS$_p \times \bS^q$ limits of brane dynamics, where the brane charge is carried by antisymmetric tensor flux through the sphere. 

Tools to study these limits are typically restricted to analysis in
effective field theory, due to technical difficulties in dealing with
string theory in backgrounds involving fluxes of Ramond antisymmetric
tensor fields.  An exception to this rule is string theory on
AdS$_3\times \bS^3\times \cM$, where $\cM=\bT^4$ or $K3$.  In this
instance, there is a duality frame where the background charges
correspond to Neveu-Schwarz (NS5) fivebranes and fundamental (F1)
strings, which are carried by NS 3-form flux $H_3$ on $\bS^3$ and
AdS$_3$, respectively.  The worldsheet CFT describing string
propagation on AdS$_3\times \bS^3\times \cM$ is exactly solvable,%
\footnote{When $\cM=K3$, at special points in the $K3$ moduli space.}
owing to its extensive current algebra symmetries; thus one can access stringy properties of the background, at least in perturbation theory around the vacuum.

String theory on AdS$_3$ is holographically dual to a two-dimensional
spacetime CFT characterized by the background charge quanta $n_1$ and $n_5$.%
\footnote{We will take pains to distinguish this two-dimensional {\it spacetime} CFT from the two-dimensional {\it worldsheet} CFT that governs first-quantized string propagation.}
The Ramond ground states of the spacetime CFT preserve half of its supersymmetries.
On the supergravity side of the duality, these states comprise a collection of NS5-F1 {\it supertubes}~\cite{Balasubramanian:2000rt,Maldacena:2000dr,Mateos:2001qs,Lunin:2001fv,Lunin:2002iz,Kanitscheider:2007wq}.  These supertubes are labelled by a partition $\{N_k^{(s)}\}$ of $N=n_1n_5$, where $s$ labels any of 8 bosonic and 8 fermionic polarization states, such that
\be
\sum_{k,s} kN_k^{(s)} = N \,.
\ee
There is a macroscopic number of such supertubes; their entropy is
\be
S_{\rm supertube} =  2\pi\sqrt{{\hat c}_\eff(n_1n_5-|\stJ_L|)}
\ee
where $\stJ_L$ is the left-moving $\bS^3$ angular momentum, and $\hat c_\eff=2,4$ for $\bT^4$ and $K3$ respectively.

Each supertube is a distinct background for string propagation; the background is semi-classical when the $N_k^{(s)}$ are macroscopic, and smooth modulo potential orbifold singularities~\cite{Lunin:2001fv,Lunin:2002iz,Kanitscheider:2007wq}.  Standard examples considered in the literature are the supertubes with $N_k^{(s)}$ nonzero only for a single value of $k$ and a single polarization $s=(++)$; these examples are particularly tractable because they preserve a great deal of symmetry~\cite{Balasubramanian:2000rt,Maldacena:2000dr,Lunin:2001jy}.  
Recent work~\cite{Martinec:2017ztd} has shown that these supertubes have an exact worldsheet description as a tensor product of Wess-Zumino-Witten (WZW) models based on the group
\be
\mathcal{G} =
\bR_t\times\bS^1_y\times SL(2,\mathbb{R}) \times SU(2) \times \bT^4
\ee
where particular left and right null currents are gauged.

This development opens up the possibility of studying stringy effects in backgrounds that are not close to the AdS vacuum.  Indeed, in the limit $k\to N$, the supertube background approaches an extremal BTZ black hole with vanishing AdS angular momentum.  The gauged WZW construction exhibits a variety of stringy effects, including a candidate for the {\it ``long string sector''} of the spacetime CFT, here described in a formalism which incorporates low-energy effective supergravity.

Additional backgrounds are obtained by spectral flow~\cite{Schwimmer:1986mf} in the $\cN=(4,4)$ superconformal representation theory of the spacetime CFT.  Spectral flow by an amount~$\alpha$ maps a state of conformal dimensions $(\stL_0,\bar \stL_0)$ and $SU(2)$ $\cR$-charges $(\stJ_L,\stJ_R)$ to a state with quantum numbers shifted by 
\be
\label{STspecflow}
(\delta \stJ_L,\delta \stJ_R) = N(\alpha,\bar\alpha) \;,~~~~
(\delta \stL_0,\delta\bar \stL_0) = N(\alpha^2,\bar\alpha^2) + 2(\alpha \stJ_L, \bar\alpha \stJ_R) \;.
\ee
When the spectral flow parameter $\alpha$ is an integer, the map takes a non-trivial chiral primary state to a superconformal descendant of that state. 
Spectral flow by $\alpha=1$ maps the vacuum state to the maximally spinning chiral primary.
When $\alpha$ is an integer plus one-half, spectral flow takes NS sector states to R sector states, and vice versa. Spectral flow by $(\alpha,\bar\alpha) = (1/2,1/2)$ takes chiral primary states in the NS-NS sector to 1/2-BPS ground states in the R-R sector. 
Spectral flow by (independent) integer amounts $\alpha \in \bZ$, $\bar\alpha\in \bZ$ is an automorphism of the CFT spectrum. 

In~\cite{Giusto:2012yz}, it was shown that for the special single-mode supertubes above, particular fractional values of $\alpha$ and $\bar\alpha$ also make sense, due to the additional symmetry present; see also~\cite{Martinec:2001cf,Martinec:2002xq,Avery:2009xr}.  The fractionally flowed states are non-trivial excited states carrying angular momenta $\stJ_L, \stJ_R$ as well as momentum charge $P=\stL_0-\bar \stL_0$. 
The general set of states with both non-zero left and right fractional spectral flow corresponds to the AdS decoupling limit of the family of non-supersymmetric smooth horizonless supergravity solutions found by Jejjala, Madden, Ross and Titchener (JMaRT)~\cite{Jejjala:2005yu}, as was shown in~\cite{Chakrabarty:2015foa}, building on~\cite{Chowdhury:2007jx,Avery:2009xr}.  
When the non-trivial spectral flow is in only one sector (say the
left-moving sector), and the other sector is in a Ramond ground state,
the backgrounds reduce to the supersymmetric solutions studied in~\cite{Giusto:2012yz} and first constructed in~\cite{Giusto:2004id,Giusto:2004ip,Jejjala:2005yu}.
The full family of these backgrounds can also be described in the gauged WZW framework, by suitably modifying the embedding of the subgroup $U(1)_L\times U(1)_R\subset\cG$ being gauged~\cite{Martinec:2017ztd}.

The JMaRT solutions are the original example of non-supersymmetric microstate solutions, and their rich physics in the supergravity sector has been well-studied~\cite{Cardoso:2005gj,Chowdhury:2007jx,Chowdhury:2008bd,Chowdhury:2008uj,Avery:2009xr,Chakrabarty:2015foa}, as we will review in due course. In addition, recent progress has been made on constructing more general families of non-supersymmetric microstate solutions that contain the JMaRT solutions~\cite{Bossard:2014ola,Bena:2015drs,Bena:2016dbw,Bossard:2017vii}.

In this paper we compute and analyze the closed string spectrum in the JMaRT backgrounds, in the NS5-brane decoupling limit.  We begin in Section~\ref{sec:sugrasolns} with a review of the asymptotically flat JMaRT solutions, from which all the spectrally-flowed states described above can be realized by specialization of the parameters.  The NS5-brane decoupling limit leads to an asymptotically linear dilaton geometry; a further F1 decoupling limit yields asymptotically AdS$_3$ spacetimes.

In Section~\ref{sec:WS CFT} we introduce the gauged WZW model which is the heart of the construction.  We review how the JMaRT solution (in the NS5-brane decoupling limit) arises after integrating out the gauge field, and review the map between the embedding parameters of the gauge group and the parameters of the supergravity solution.  The gauged WZW model thus describes an asymptotically linear dilaton spacetime, dual to little string theory (LST), with an AdS$_3$ cap in the core of the solution.

The perturbative string spectrum is analyzed in Section~\ref{sec:Spectrum}, beginning with low-lying supergravity states bound to the cap, and scattering states that are plane waves in the linear dilaton region.  We then discuss the effect of {\it worldsheet} spectral flow in the $SU(2)$ and $SL(2,\bR)$ components of the WZW model on $\cG$, followed by the structure and spectrum of strings winding $\bS^1_y$.  The spectrally flowed strings typically have a large amount of angular momentum on $\bS^3$ and/or AdS$_3$ which causes them to expand as a result of their interaction with the background flux --  they are the analogues of giant gravitons~\cite{McGreevy:2000cw,Lunin:2002bj}.  

A key feature is the interplay of worldsheet spectral flow and large gauge transformations.  Large gauge transformations relate the winding number on $\bS^1_y$ to the spectral flow parameters, trading one for another, and since the amount of spectral flow is not conserved in correlation functions, neither is winding on $\bS^1_y$.  However, strings winding $\bS^1_y$ carry the same F1 charge as the background electric $H_3$ flux, and the total such F1 charge is conserved.  We thus arrive at a physical interpretation of large gauge transformations as implementing brane/flux transitions, wherein F1 charge (as well as momentum and angular momentum) is exchanged between the background and the set of perturbative strings in it.
 
Finally, in Section~\ref{sec:Correlators} we elaborate on the structure of the two-point amplitude in this family of string backgrounds, following~\cite{Giveon:2015cma,Martinec:2017ztd}.   Since the geometry involves the decoupled fivebrane throat of little string theory, not surprisingly the perturbative string S-matrix has a structure similar to that of NS fivebranes on their Coulomb branch explored in~\cite{Aharony:2004xn} in the context of LST.  The two-point function is the reflection amplitude for scattering strings off the cap of the geometry.  The standard relation between poles in the reflection amplitude and the bound state spectrum is borne out -- the two-point function exhibits a series of poles associated to the spectrum of supergravity solutions bound to the cap worked out in Section~\ref{sec:Spectrum}.  An additional, stringy set of poles is given a physical interpretation in terms of the sub-string-scale microstructure of the fivebrane sources in the background, adapting the analysis of~\cite{Giveon:2015cma} to reveal properties of the cap invisible to low-energy supergravity probes.

A recurring thread in the discussion concerns potential instabilities
of the background.  The non-supersymmetric asymptotically flat JMaRT geometries have 
an ergoregion for any choice of asymptotically timelike Killing vector field~\cite{Jejjala:2005yu}, as
we review in Appendix~\ref{app:ergoregion}. There is a corresponding ergoregion instability, 
as we review in Section~\ref{sec:supergravity}.
In contrast, the supersymmetric backgrounds of \cite{Giusto:2004id,Giusto:2004ip,Jejjala:2005yu,Giusto:2012yz} have a globally null Killing vector field (arising from supersymmetry~\cite{Gutowski:2003rg}) and are expected to be linearly stable in supergravity, while the question of non-linear stability has recently received interest~\cite{Eperon:2016cdd,Marolf:2016nwu,Bena:2017xbt}.  In the fivebrane
decoupling limit, the imaginary parts of the frequencies of the unstable modes 
tend to zero, and a globally timelike Killing vector field appears due to the modified 
asymptotics of the decoupled throat geometry. However
one can identify the ingredients of the instability of the corresponding asymptotically flat solutions 
through an analysis of the local
energy eigenvalues at the top and the bottom of the throat.  In the
null-gauged WZW model, the gauge orbits in the target-space group manifold $\cG$ are mostly along
$\bR_t \times\bS_y^1$ at the bottom of the throat, while they are mostly along $\sltwo$ asymptotically, so that the physical local energy in the cap is the $\sltwo$ energy, while asymptotically it is the $\bR_t$ momentum.  One finds that for the incipient unstable modes, these two timelike momenta indeed have opposite signs.

One might also expect that the non-supersymmetric backgrounds exhibit
stringy instabilities, which in addition might not shut off in the
decoupling limit.  The effective geometry in the AdS$_3$ decoupling
limit of the JMaRT solution is an orbifold $(\ads3 \times \bS^3)/\bZ_k$~\cite{Jejjala:2005yu,Chakrabarty:2015foa}.  Generically, the orbifold singularities of these solutions are non-supersymmetric Hirzebruch-Jung singularities. Solutions with such non-supersymmetric orbifold singularities typically have tachyons in twisted sectors; moreover, the presence of such a tachyon should not depend on the asymptotic structure of the geometry -- only the vicinity of the orbifold point should matter.  Nevertheless, we find that there are no tachyonic solutions to the physical state constraints, even in the winding sectors that are the analogues of orbifold twist sectors in the null gauging approach.
As a byproduct of the analysis we clarify that, while the effective geometry is an orbifold geometry, the worldsheet theory is {\it not} an orbifold CFT, so results from string theory on non-supersymmetric orbifolds do not directly apply. Amongst the rich spectrum of both supergravity and excited string states, we find no unstable modes in the NS5-brane decoupling limit.

\section{Supergravity solutions}
\label{sec:sugrasolns}

The JMaRT solutions were obtained in~\cite{Jejjala:2005yu} by imposing 
smoothness and absence of horizons on the general family of non-extremal three-charge
solutions of~\cite{Cvetic:1996xz,Cvetic:1997uw}. 
The general holographic
description of the AdS decoupling limit of these solutions was identified in~\cite{Chakrabarty:2015foa}. In
this section we briefly review the analysis of~\cite{Chakrabarty:2015foa} and we derive some
relations that will be useful in what follows.

\subsection{The JMaRT solutions}

We write the JMaRT solutions in the NS5-F1-P frame, 
S-dual to the D1-D5-P frame discussed in~\cite{Chakrabarty:2015foa}. In the next
section we will realize the NS5 decoupling limit of these solutions as
the target-space geometry of a gauged WZW model. The string-frame
metric is given by:
\begin{align}
ds^2  \,=&~\, \frac{f_0}{\tilde H_1} (- dt^2 + dy^2) + \frac{M}{\tilde H_1}(c_p dt -
  s_p dy)^2 \nonumber \\
& +\tilde H_5 \Big(\frac{r^2 dr^2}{(r^2+a_1^2)(r^2+a_2^2)-M r^2}+
  d\theta^2\Big) \nn\\
& + \Big( \tilde H_5 - (a_2^2-a_1^2) \frac{(\tilde H_1 +\tilde
  H_5-f_0)\cos^2\theta}{\tilde H_1}\Big)\cos^2\theta d\psi^2  \nonumber \\
& + \Big( \tilde H_5 + (a_2^2-a_1^2) \frac{(\tilde H_1 +\tilde
  H_5-f_0)\sin^2\theta}{\tilde H_1}\Big)\sin^2\theta d\phi^2  \label{metricJMaRT} \\
& + \frac{M}{\tilde H_1} (a_1 \cos^2\theta d\psi + a_2 \sin^2 \theta d\phi)^2
  \nn\\
& + \frac{2M\cos^2\theta}{\tilde H_1} \Big[ (a_1 c_1 c_5 c_p - a_2 s_1
  s_5 s_p)dt + (a_2s_1s_5c_p- a_1 c_1 c_5 s_p) dy\Big] d\psi
  \nonumber\\
& + \frac{2M\sin^2\theta}{\tilde H_1} \Big[ (a_2 c_1 c_5 c_p - a_1 s_1
  s_5 s_p)dt + (a_1s_1s_5c_p- a_2 c_1 c_5 s_p) dy\Big] d\phi
  \nonumber\\
& +\sum_{a=1}^4 dz_a^2 \, , \nn
\end{align}
where we take the $y$~circle to have radius $R$ at spacelike infinity, $y \sim y + 2 \pi R$, and
\begin{align}
\tilde H_i & = f_0 +M s_i^2 \, , \quad f_0 = r^2 +
  a_1^2\sin^2\theta + a_2^2 \cos^2\theta \, , \nonumber\\
c_i &= \cosh \delta_i \, ,\quad s_i  = \sinh \delta_i \, .
\end{align}
The B-field is
\begin{align}
B_2 \,=&~\, \frac{M \cos^2\! \theta }{\tilde H_1}\Big[(a_2 c_1 s_5 c_p -
      a_1 s_1 c_5 s_p)
      dt+ (a_1 s_1 c_5 c_p - a_2 c_1 s_5 s_p)dy\Big]\wedge
      d\psi \nn\\
& + \frac{M \sin^2\!\theta}{\tilde H_1}\Big[ (a_1 c_1 s_5 c_p - a_2
  s_1 c_5 s_p)
      dt+ (a_2 s_1 c_5 c_p - a_1 c_1 s_5 s_p)dy \Big]\wedge d \phi  \, \label{BfieldJMaRT}\\
& -\frac{M s_1 c_1}{\tilde H_1} dt \wedge dy - \frac{M s_5 c_5}{\tilde
  H_1} \Big(
 r^2 +a_2^2 + M s_1^2\Big)
  \cos^2\!\theta d\psi \wedge d\phi \, , \nn
\end{align}
and the dilaton is given by
\begin{equation}
e^{2\Phi} = \frac{\tilde H_5}{\tilde H_1} \, .
\end{equation}
The charges $Q_i$ are given by 
\be \label{eq:charges}
Q_i = M s_i c_i \,, \qquad i=5,1,p \,.
\ee
The ADM mass and angular momenta are, in units in which $4 G^{(5)}/\pi =
1$,
\begin{align}
M_{ADM} &= \frac{M}{2}(c_1^2+c_5^2+ c_p^2 + s_1^2+s_5^2+s_p^2) \, , \label{JMaRTADM}\\
J_{\psi} &= - M (a_1 c_1 c_5 c_p - a_2 s_1 s_5 s_p) \, , \\
J_{\phi} &= - M (a_2 c_1 c_5 c_p - a_1 s_1 s_5 s_p) \, .
\end{align} 
To have positive mass we take $M \geq 0$ and without loss of
generality we take $\delta_1, \delta_5, \delta_p \geq 0$ and $a_1 \geq
a_2 \geq 0$.
In order to have smooth horizonless solutions, one must investigate
the behavior of the metric at the roots of the function $g(r) =
(r^2+a_1^2)(r^2+a_2^2)-M r^2$, given by
\begin{equation}\label{defrpm}
r^2_{\pm} = \frac12\left[ M-a_1^2-a_2^2 \pm \sqrt{(M-a_1^2
    -a_2^2)^2-4a_1^2a_2^2}\right] \, .
\end{equation} 
As discussed in detail in~\cite{Jejjala:2005yu,Chakrabarty:2015foa},
this analysis imposes the following relations among the parameters of the
solution:
\begin{align}
a_1 a_2 & = \frac{Q_1 Q_5}{(k R)^2}\frac{s_1^2c_1^2 s_5^2 c_5^2 s_p
          c_p}{ (c_1^2 c_5^2 c_p^2 - s_1^2 s_5^2 s_p^2)^2} \, , \label{eq:a1a2}\\
M  & = a_1^2 + a_2^2 - a_1 a_2 \frac{(c_1^2 c_5^2 c_p^2 +
  s_1^2s_5^2s_p^2)}{s_1 c_1 s_5 c_5 s_p c_p}\, , \label{eq:M}\\
\n  & = (kR)\frac{s_p c_p}{(a_1 c_1 c_5 c_p - a_2 s_1 s_5 s_p)} \in \mathbb{Z}\, , \label{eq:n}\\
\m & = -(kR) \frac{s_p c_p}{(a_2 c_1 c_5 c_p - a_1 s_1 s_5 s_p)}\in \mathbb{Z} \, . \label{eq:m}
\end{align}
It is useful to introduce the quantities
\begin{equation}
\mathfrak{j} = \sqrt{\frac{a_2}{a_1}} \, ,\quad \mathfrak{s} =\sqrt{
\frac{s_1s_5s_p}{c_1c_5c_p} }\, , \quad \mathfrak{j}, \mathfrak{s} \leq 1
\, .
\end{equation}
We have the relations
\begin{equation}\label{eq:mn}
\m + \n \,=\, \frac{\mathfrak{j} - \mathfrak{j}^{-1}}{\mathfrak{s}-\mathfrak{s}^{-1}} \,\equiv\, 2s+1 \,,
\qquad \m - \n \,=\,
\frac{\mathfrak{j} + \mathfrak{j}^{-1}}{\mathfrak{s}+\mathfrak{s}^{-1}} \,\equiv\, 2\bar s+1  \,,
\end{equation}
where the integers $s,\bar s$ are introduced for future reference.
The roots of $g(r)$ in \eqref{defrpm} are given by
\begin{equation} \label{eq:rp-rm-s}
r_{+}^2 = - a_1 a_2 \mathfrak{s}^2 \, , \quad r_{-}^2 = - a_1 a_2
\mathfrak{s}^{-2} \, .
\end{equation}
By using \eqref{eq:a1a2} one can write $M$ in the form
\begin{equation}\label{MQ1Q5}
M = a_1 a_2(\mathfrak{s}^2-\mathfrak{j}^2)(\mathfrak{j}^{-2}\mathfrak{s}^{-2}-1) =a_1 a_2 \m \n (\mathfrak{s}^2 -\mathfrak{s}^{-2})^2 =  \frac{Q_1 Q_5 \m \n}{(k R)^2 c_p s_p}\, .
\end{equation}
From the condition $M\geq 0$ we see that $\mathfrak{s}^2>
\mathfrak{j}^2$ and, from \eqref{eq:mn}, that $\m>\n$.
Note that $c_p s_p = Q_p/M$, so that
\begin{equation}\label{eq:Qpmn}
Q_p = \frac{Q_1 Q_5}{(k R)^2} \m \n \, .
\end{equation}
From \eq{eq:n}, \eq{eq:m}, and \eqref{eq:Qpmn}, we then obtain the angular momenta
\be \label{eq:JpsiJphi-2}
J_{\psi} = - \m \frac{Q_1 Q_5}{k R} \,, \qquad
J_{\phi} = \,\n \frac{Q_1 Q_5}{k R} \, .
\ee

\subsection{Fivebrane decoupling limit}
\label{sec:fivebranedecouplinglimit}

We are principally interested in the fivebrane decoupling limit of the above geometry, where one takes $\delta_5\to\infty$ with $Me^{2\delta_5}$ held fixed.  
In this limit, an exact worldsheet description of string propagation is available~\cite{Martinec:2017ztd}.
There is a further limit that yields an asymptotically AdS$_3
\times \mathbb{S}^3 \times \mathbb{T}^4$ geometry,
as we will discuss shortly. 

Taking $\delta_5\rightarrow \infty$ means
that $c_5 \simeq s_5 \simeq \tha e^{\delta_5}$. 
In more detail, we have 
\be
c_5^2 \;\simeq\; s_5^2 \;\simeq\; \frac{Q_5}{M}
\ee
as $M \to 0$ with $Q_5$ fixed. From \eq{eq:charges}, \eq{eq:a1a2}, and \eq{eq:M} we have 
\be \label{eq:fivebrane-dec-limit}
Q_1 \,=\, \cO(M) \,, \qquad Q_p \,=\, \cO(M) \,, \qquad a_1 \,=\,\cO(M^{1/2}) \,, \qquad  a_2 \,=\,\cO(M^{1/2})
\ee
and we see from \eq{eq:Qpmn} that $R$ is order $M^0$. So we have a hierarchy between the larger scales $Q_5$, $R$ and the other scales.

Restricting to a region $r^2 \ll Q_5$, in the above limit
we can approximate the five-brane harmonic function as 
$\tilde H_5 \approx Q_5$.  One can also take this limit by replacing
\be \label{eq:NS5-scal-limit}
r \to \epsilon \;\! r \,, \qquad M \to \epsilon^2 M \,, \qquad a_1\to \epsilon \;\! a_1 \,, \qquad a_2 \to \epsilon \;\! a_2
\ee
and taking $\epsilon \to 0$.
In this limit we obtain:
\begin{eqnarray}
ds^2 \,&=&\, \frac{f_0}{\Sigma}\bigl(-\,dt^2+dy^2\bigr) +\frac{M}{\Sigma}\bigl( c_p \,dt - s_p \,dy\bigr)^2
+ Q_5 \bigl( d\rho^2+d\theta^2\bigr)
\nn\\
&&
+\frac{Q_5}{\Sigma}\Big[ \bigl(r_+^2-r_-^2\bigr)\cosh^2\rho + r_-^2 +
  a_2^2 + Ms_1^2\Big]\sin^2\theta\, d\phi^2 \nn
\\
&&
+\frac{Q_5}{\Sigma}\Big[ \left(r_+^2-r_-^2\right)\sinh^2\rho + r_+^2 + a_1^2 + M s_1^2\Big]\cos^2\theta\, d\psi^2
\label{eq:CFTmetric}\\
&&
+\frac{2\sqrt{M Q_5}\cos^2\theta}{\Sigma}\Big[ \bigl(a_1c_1c_p-a_2s_1s_p\bigr) \, dt+ \bigl(a_2s_1c_p-a_1c_1s_p\bigr)\, dy\Big]\,d\psi
\nn\\
&&
+\frac{2\sqrt{M Q_5}\sin^2\theta}{\Sigma}\Big[ \bigl(a_2c_1c_p-a_1s_1s_p\bigr) \, dt+ \bigl(a_1s_1c_p-a_2c_1s_p\bigr)\, dy\Big]\,d\phi
+\sum_{a=1}^4 dz_a^2 \,,
\nn \\[.3cm]
B_2 \,&=&\, \frac{\sqrt{M Q_5}}{\Sigma}\cos^2\theta\Big[(a_2 c_1 c_p - a_1 s_1 s_p)
      dt+ (a_1 s_1 c_p -a_2 c_1 s_p)dy\Big]\wedge
      d\psi  \cr
&& + \frac{\sqrt{M Q_5}}{\Sigma}\sin^2\theta\Big[ (a_1 c_1 c_p - a_2 s_1 s_p)
      dt+ (a_2 s_1 c_p - a_1 c_1 s_p)dy \Big]\wedge d \phi  \label{eq:CFTBfield}\\
&& -\frac{M s_1 c_1}{\Sigma} dt \wedge dy - \frac{Q_5}{\Sigma} \Big[
  (r_{+}^2-r_{-}^2)\sinh^2\!\rho + r_{+}^2 +a_2^2 + M s_1^2\Big]
  \cos^2\!\theta \, d\psi \wedge d\phi \,,  \nn \\[.3cm]
e^{2\Phi} \,&=&\, \frac{g_s^2 Q_5}{\Sigma} \,,\vspace{-4mm} \label{eq:CFT-dil}
\end{eqnarray}
where we introduced a new radial coordinate
\begin{equation}\label{defrho}
\sinh^2 \! \rho = \frac{r^2-r_+^2}{r_+^2-r_-^2} \, ,
\end{equation}
and we defined
\begin{equation}
\Sigma = \tilde H_1 =  f_0 + M s_1^2 =\frac12\left[(r_+^2-r_-^2)\cosh 2\rho +(a_2^2-a_1^2) \cos
  2\theta + M \right] + M s_1^2  \, .
\end{equation}
The last relation follows by taking the $\delta_5 \rightarrow \infty$
limit in \eqref{eq:M} and \eq{eq:rp-rm-s}. In this limit, the ADM mass
of the full solution \eq{JMaRTADM} becomes
\begin{equation}\label{eq:NS5decoupledMadm}
M_{ADM} = Q_5 + M(c_1^2c_p^2-s_1^2s_p^2) \,,
\end{equation}
where the second term on the right-hand side is a subleading correction to the first.

The asymptotically flat region has been decoupled from the
geometry and the asymptotics are now the linear dilaton throat of the
fivebranes, 
\begin{align}
ds^2 & \;\sim\; -dt^2 + dy^2 + Q_5 (d\rho^2 + d\Omega_3^2 ) +\sum_{a=1}^4 dz_a^2 \, , \\
\Phi & \;\sim\; -\rho \, .
\end{align}
In Section \ref{sec:WS CFT} we will construct an exact worldsheet CFT for the solution \eqref{eq:CFTmetric}, \eqref{eq:CFTBfield}.

\subsection{AdS decoupling limit}
\label{sec:AdS}

The further AdS$_3 \times \mathbb{S}^3 \times \mathbb{T}^4$ limit is 
obtained by simultaneously sending $R\to \infty$ with scaled
energies $ER$ and $y$-momenta $P_y R$ held fixed, and sending $\delta_1 \rightarrow \infty$ with $Me^{2\delta_1}$ held fixed, in addition to the fivebrane limit taken above ($\delta_5 \rightarrow \infty$ with $Me^{2\delta_5}$ held fixed).
This means that as $M \to 0$ we have
\be
Q_5 \,=\, \cO(M^0) \,, \quad~~ Q_1 \,=\, \cO(M^0) \,, \quad~~ Q_p \,=\, \cO(M) \,, \quad~~ a_1 \,, a_2 \,=\,\cO(M^{1/2}) \,,
\ee
and so from \eq{eq:Qpmn} we see that
\be
R \,=\, \cO(M^{-1/2}) \,
\ee
is indeed now the largest lengthscale in the problem.

In this limit one can express the F1 charge as $Q_1 \simeq \tfrac{1}{4} M e^{2\delta_1}$, and we have
\be \label{eq:c1s1-ads-lim}
c_1^2 \;\simeq\; s_1^2 \;\simeq\; \frac{Q_1}{M} \,.
\ee
The AdS throat is then the region $r^2 \ll Q_1 $, where one can approximate $\tilde{H}_1 = \Sigma \simeq Q_1$.
One can take this limit by defining 
\be \label{eq:t-y-tilde}
\tilde{t} = \frac{t}{R} \,, \qquad \tilde{y} = \frac{y}{R} \,, 
\ee
making the replacements
\be \label{eq:ads-scal-lim}
r \to \epsilon \;\! r \,,\qquad M \to \epsilon^2 M \,, \qquad a_1\to \epsilon \;\! a_1 \,, \qquad a_2 \to \epsilon \;\! a_2 \,, \qquad R \to \frac{R}{\epsilon}
\ee
and then taking $\epsilon \to 0$ at fixed $\tilde{t}$, $\tilde{y}$. Note that while \eq{eq:ads-scal-lim} may seem similar to \eq{eq:NS5-scal-limit}, the scaling of $c_1, s_1$ in \eq{eq:c1s1-ads-lim} means that it is quite different; in particular now $\tilde{H}_1 = \Sigma$ becomes $Q_1$, which stays finite, rather than scaling as $\epsilon^2$.

One then obtains an asymptotically AdS$_3 \times \mathbb{S}^3 \times
\mathbb{T}^4$ solution.  In this limit the six-dimensional part of the metric
is given by
\begin{align} \label{eq:ads}
\frac{1}{Q_5}\;\!ds_6^2 \,=&~ -\frac{1}{k^2}\cosh^2 \! \rho \, d
                           \tilde t^2 +
           d \rho^2 +
    \frac{1}{k^2}   \sinh^2\!  \rho\, d \tilde y^2 \\ &~ +
d\theta^2 + \sin^2\! \theta \left[ d\phi + \frac{\m}{k} d \tilde y
  -\frac{\n}{k} d \tilde t\right]^2 + \cos^2\! \theta \left[ d\psi
  -\frac{\n}{k}d \tilde y + \frac{\m}{k} d \tilde t\right]^2 . \nn
\end{align}
In this limit the ADM mass of the full solution \eq{JMaRTADM} becomes
\begin{equation}\label{ADMmasslargeR}
M_{ADM} = Q_5 + Q_1 + \frac{Q_1 Q_5}{R^2} \frac{\m^2 + \n^2 -1}{2
  k^2} \,
\end{equation}
where on the right-hand side there is now a double hierarchy: of the three terms, the third term is a subleading correction to the second term, which is itself a subleading correction to the first term.

This AdS$_3 \times \mathbb{S}^3 \times \mathbb{T}^4$ throat has an
interesting orbifold structure depending on the parameters $(k,\m,\n)$,
as we now describe.

\subsection{Supersymmetric and non-supersymmetric \texorpdfstring{AdS$_3 \times $S$^3$}{} orbifolds}
\label{OrbifoldStructure}

We now analyze the structure of orbifold singularities in the core of the solutions following~\cite{Jejjala:2005yu,Chakrabarty:2015foa}. The analysis can be performed on any of the asymptotically flat, asymptotically linear dilaton, or asymptotically AdS solutions, with the same results. For ease of presentation we shall describe the analysis of the asymptotically AdS$_3 \times \mathbb{S}^3 \times \mathbb{T}^4$ solutions.

It is useful to introduce new coordinates
\begin{equation}\label{AdS3variables}
 \tilde{\psi} = \psi
  -\frac{\n}{k} \tilde y + \frac{\m}{k} \tilde t \, , \qquad
  \tilde{\phi} = \phi + \frac{\m}{k} \tilde y
  -\frac{\n}{k} \tilde t \, ,
\end{equation}
in terms of which the six-dimensional metric \eq{eq:ads} takes the standard global AdS$_3 \times \bS^3$ form. The relations \eq{AdS3variables} are interpreted as bulk spectral flow~\cite{Giusto:2004id,Bena:2008wt,Giusto:2012yz,Bena:2017geu}. The periodicities of the angles at infinity ($y \sim y+2\pi R$, $\psi \sim\psi+2\pi$, $\phi \sim\phi+2\pi$) induce the following periodic identifications:
\begin{align}
\label{PrimitiveShift}
&A:  (\tilde y/k, \tilde \psi, \tilde \phi)  \sim  (\tilde y/k, \tilde \psi,
                                       \tilde \phi)+\frac{2\pi}{k} (1,
                                       -\n, \m) \, ,\\
&B:  (\tilde y, \tilde \psi, \tilde \phi) \sim (\tilde y, \tilde \psi, \tilde
                                       \phi)+  2\pi (0,1,0) \, , \nn \\
&C:  (\tilde y, \tilde \psi, \tilde \phi)  \sim (\tilde y, \tilde \psi, \tilde
                                       \phi)+  2\pi (0,0,1) \nn \, .
\end{align}
For the metric \eq{eq:ads} to describe a smooth shrinking of the $\tilde{y}$ circle at $\rho=0$, the combination $\tilde{y}/k$ should have period $2\pi$. The above lattice of identifications contains fixed points that give rise to orbifold singularities, as follows.

If $\gcd (k,\m) = \ell_1$ and $\gcd
(k,\n)=1$, with $\ell_1 >1$, there is a $\mathbb{Z}_{\ell_1}$ orbifold
singularity at $\rho = 0$, $\theta = \pi/2$ and the solution is otherwise smooth. At this locus $\tilde{y}$ and $\tilde\psi$ shrink, while $\tilde \phi$ remains finite. Setting $k = \ell_1
\hat k$, $\m = \ell_1 \hat m$, an action that leaves $\tilde \phi$
invariant is  $A^{\hat k} C^{-\hat m}$. Since $\m/\ell_1 = \hat m$ is an integer, it is equivalent to consider
\begin{equation}
\quad A^{\hat k} B^{\hat m} C^{-\hat m} ~:~~ 
\left(\frac{\tilde y}{k}, \tilde \psi, \tilde \phi\right)\sim
\left(\frac{\tilde y}{k}, \tilde \psi, \tilde \phi\right)+ 2\pi \left(\frac{1}{\ell_1}, \frac{\m-\n}{\ell_1}, 0
\right) . \label{eq:orb-case-1}
\end{equation}
Defining $\tilde \theta = \theta - \pi/2$, $Z_1 = \rho e^{i \tilde
  y/k}$, $Z_2 = \tilde \theta e^{i \tilde \psi}$, locally the orbifold
action on the $(Z_1, Z_2)$ plane is of the form
\begin{equation}
A^{\hat k} B^{\hat m}  C^{-\hat m}(Z_1, Z_2 ) \;=\; (\omega Z_1 , \omega^{p} Z_2) \,
,\quad~ \omega = e^{2\pi i/\ell_1} \, ,\quad~ p = \m-\n = 2\bar s+1 \, .
\end{equation}
This is a Hirzebruch-Jung singularity, of the kind
discussed in~\cite{Adams:2001sv,Harvey:2001wm}. Following these references we denote the orbifold singularity by $\mathbb{C}^2/\mathbb{Z}_{\ell_1 (p)}$.  For $p = 1$ and with the appropriate action of the rotation on spacetime fermions, the orbifold is a standard supersymmetric $A_{(\ell_1-1)}$-type
singularity. This is consistent with the fact that the JMaRT solutions
with $\m=\n+1$ correspond to the supersymmetric three-charge solutions
discussed in \cite{Giusto:2012yz}, with right moving spectral flow
parameter $\bar s = 0$. For $\bar s >0$ we have
$p>1$ and the orbifold generically breaks all the supersymmetries. 

A similar analysis applies for the case $\gcd (k,\n) = \ell_2>1$ and $\gcd
(k,\m)=1$. Now there is a singularity at $\rho=0$, $\theta = 0$ and the solution is otherwise smooth. At
this point $\tilde \psi$ remains finite while $\tilde \phi$
shrinks. Defining $k = \ell_2 \hat k$, $\n =\ell_2 \hat n$, an identification that leaves $\tilde \psi$ invariant is
\begin{equation}
A^{\hat k} B^{\hat n} C^{-\hat n} : 
\left(\frac{\tilde y}{k}, \tilde \psi, \tilde \phi\right)\sim
\left(\frac{\tilde y}{k}, \tilde \psi, \tilde \phi\right)+ 2\pi
\left(\frac{1}{\ell_2},0, \frac{\m-\n}{\ell_2}
\right) \, ,
\end{equation}
and we obtain a $\mathbb{C}^2/\mathbb{Z}_{\ell_2(p)}$ singularity
acting on  $Z_1 = \rho e^{i \tilde
  y/k}$, $Z'_2 = \theta e^{i \tilde\phi}$. For $p=1$, or
equivalently $\bar
s = 0$, this is the supersymmetric orbifold discussed in
\cite{Giusto:2012yz}, while generically the orbifold action is
incompatible with supersymmetry.

When $\gcd (k,\m) = \ell_1>1$ and $\gcd(k,\n)=\ell_2>1$, there are two sub-cases. First, when $\gcd(k,\m,\n) = 1$, from the analysis above there is a $\mathbb{C}^2/\mathbb{Z}_{\ell_1(p)}$ singularity at $\rho = 0$, $\theta =\pi/2$, a $\mathbb{C}^2/\mathbb{Z}_{\ell_2(p)}$ singularity at $\rho = 0$, $\theta =0$, and the solution is otherwise smooth.
Second, when $\gcd(k,\m,\n) = \ell_3 >1$, in addition to the above
singularities, there is now a $\mathbb{C}/\mathbb{Z}_{\ell_3}$ orbifold  singularity at $\rho=0$ for all values of $\theta$. This can be seen as follows. We define $k = \ell_3
\hat \ell_1 \hat \ell_2 \hat k$, $\m = \ell_3 \hat \ell_1 \hat m$, $\n = \ell_3 \hat
\ell_2 \hat n$. For $0<\theta<\pi/2$ both $\tilde \psi$ and $\tilde
\phi$ are of finite size. The action that leaves them invariant is
\begin{equation}
\quad A^{\hat \ell_1 \hat \ell_2 \hat k}B^{\hat \ell_2 \hat n} C^{-\hat \ell_1 \hat
  m}~:~~ \left(\frac{\tilde y}{k}, \tilde \psi, \tilde \phi\right)\sim
\left(\frac{\tilde y}{k}, \tilde \psi, \tilde \phi\right)+ 2\pi
\left(\frac{1}{\ell_3},0, 0
\right) \, .
\end{equation}
This gives a $\mathbb{C}/\mathbb{Z}_{\ell_3}$ orbifold singularity at
$\rho = 0, 0< \theta < \pi/2$. Note that this
singularity is always absent in the supersymmetric solutions since in this
case ${\gcd(k,\m,\n) =1}$; a nontrivial $\mathbb{C}/\mathbb{Z}_{\ell_3}$ singularity breaks all supersymmetries.

String theory on non-supersymmetric orbifolds contains tachyons in twisted sectors.  Thus, when investigating the question of whether there is an instability in the fivebrane decoupling limit, it is not sufficient to analyze solutions to the wave equation.  In asymptotically locally flat spacetimes, the analysis of~\cite{Adams:2001sv,Harvey:2001wm} showed that condensation of these closed string tachyons causes spacetime to decay toward a supersymmetric background, as a pulse of radiation travels down the conical geometry to infinity, leaving behind flat space or a supersymmetric orbifold.  In the context of AdS$_3$, such a non-supersymmetric background is a coherent excited state of the spacetime CFT (as we review below), and if there are twisted sector tachyons then radiation from a decay via their condensation will reflect off the AdS$_3$ boundary and thermalize.
However we should emphasize that the gauged WZW model is {\it not} a global orbifold,%
\footnote{Such orbifolds were analyzed in~\cite{Martinec:2001cf}, where it was observed that $k$ must divide $\nfive$ for the worldsheet CFT to exist.  On the other hand, the GWZW models actually prefer $k$ and $\nfive$ to be relatively prime, so that the source consists of a single supertube rather than $\gcd(k,\nfive)$ of them (though of course this is not required).} 
and the possible presence and nature of ``twisted sectors'' remains an open question; this will be a focus of Section~\ref{sec:winding} below.

While we have exhibited particular non-supersymmetric orbifold structures for transformations that either keep $\tilde\phi$ or $\tilde\psi$ fixed, fundamentally it is the primitive shift $A$ of~\eqref{PrimitiveShift} that determines whether supersymmetry is preserved, since the Killing spinors square to Killing vectors that involve rotations along $\tilde \varphi_\pm =\tilde\phi\pm\tilde\psi$ leaving $\tilde\varphi_\mp$ fixed.  Rewriting $A$ as
\be
A :  \delta\bigl(\tilde y/k, \tilde \varphi_+, \tilde \varphi_- \bigr)  \sim 
\frac{2\pi}{k} \bigl(1,\m\tight-\n, \m\tight+\n \bigr) = \frac{2\pi}{k} \bigl(1,2\bar s\tight+1, 2s\tight+1 \bigr)
\ee
shows that left-moving supersymmetries are preserved for $s=0$, and right-moving supersymmetries are preserved for $\bar s=0$.

Independent of the above considerations of potential stringy instabilities, the non-supersymmetric asymptotically flat JMaRT solution has an ergoregion instability in the supergravity sector~\cite{Cardoso:2005gj}, even if $\gcd(k,\m)=\gcd(k,\n)=1$. This instability is interpreted holographically as a classically enhanced version of Hawking radiation~\cite{Chowdhury:2007jx,Chowdhury:2008uj}.
We shall discuss this in more detail when we solve for the supergravity spectrum in Section \ref{sec:supergravity}.
The decay rate scales as a positive power of $Q_1Q_5/\Ry^4$~\cite{Chowdhury:2007jx,Chowdhury:2008uj,Chakrabarty:2015foa} and thus vanishes in the fivebrane decoupling limit~\eq{eq:fivebrane-dec-limit}.
Consistent with this, we show in Appendix \ref{app:ergoregion} that in this limit, the decoupled geometry has a globally timelike Killing vector field. The vanishing of the instability may be thought of as a consequence of the modified asymptotics that allow a more general set of Killing vector fields to be timelike at infinity in the fivebrane decoupling limit.

\subsection{Holographic description}
\label{subsec:holographicdescription}

In this section we briefly review the holographic description of the general JMaRT solutions obtained in~\cite{Chakrabarty:2015foa}, building on the earlier works~\cite{Chowdhury:2007jx,Avery:2009xr}.

At a particular point in moduli space, the holographically dual CFT is an $\cN=(4,4)$ symmetric product orbifold CFT, with target space $\cM^N/S_N$, where $\cM=\bT^4$ or $K3$, $N=n_1 n_5$, and $S_N$ is the symmetric group. For recent discussions see e.g.~\cite{Bena:2017xbt,Bena:2016agb}.

The symmetric product orbifold CFT contains (spin-)twist operators, that link together copies of the CFT by twisting together the boundary conditions of the fields on different copies (see e.g.~\cite{Lunin:2000yv,Lunin:2001pw}). Twist operators are labelled by permutations. A  twist operator labelled by a cyclic permutation of length $k$ combines $k$ copies of the CFT with target space $\cM$ into a CFT that effectively lives on a base space circle that is $k$ times longer than the original CFT base space circle. Such a sector of the full CFT is often referred to as a `strand' of length $k$. An interesting family of states comprises those in which the $N$ copies of the CFT with target space $\cM$ are twisted into $N/k$ strands of length $k$. 

On a strand of length $k$, the conformal transformation $z=t^k$ maps to the $k$-fold covering space of the CFT's base space. For many purposes it is convenient to work in the covering space and then return to the original base space; see e.g.~\cite{Carson:2014yxa,Carson:2014xwa,Carson:2014ena,Carson:2016uwf}. The state of lowest conformal dimension on a strand of length $k$ is that for which the covering space state is the NS-NS vacuum, which we denote by $\ket{0_{\mathrm{NS}}}_k\;\!$.

On a strand of length $k$, there is an enhanced spectral flow symmetry known as fractional spectral flow~\cite{Martinec:2001cf,Martinec:2002xq,Avery:2009xr,Chakrabarty:2015foa}. This can be summarised as follows: map to the $k$-fold cover, perform spectral flow by an amount $s$, and map back to the base space. The result is an effective spectral flow by an amount $\alpha=s/k$. There are independent spectral flow operations in left and right moving sectors. 

The CFT states dual to the general JMaRT solutions are composed of $N/k$ identical strands of length $k$. The state on each strand is specified as follows. Consider the state $\ket{0_{\mathrm{NS}}}_k$, and perform fractional spectral flow with parameters $(\alpha, \bar\alpha)$ given by~\cite{Chakrabarty:2015foa}:
\be
 \alpha \;=\; \frac{s+1/2}{k} \,, \qquad \bar\alpha \;=\; \frac{\bar s+1/2}{k} \,.
\ee
Then the parameter match to the JMaRT solutions is that the parameter $k$ is the same, and the spectral flow integers $s,\bar s$ are as given in~\eqref{eq:mn},
\begin{equation}\label{mnssbar}
\m + \n = 2s +1 \, ,\qquad \m - \n = 2\bar s +1 \, .
\end{equation}
The quantum numbers of the resulting CFT states are~\cite{Chakrabarty:2015foa}:
\begin{align}
\label{JLandJR}
h = \frac{N}{4}\left[ 1+ \frac{(\m+\n)^2 - 1}{k^2}\right] \, ,\quad m_L =
  \frac{N}{2}\frac{\m + \n}{k} \, ,\\
\bar h = \frac{N}{4}\left[ 1+ \frac{(\m-\n)^2 - 1}{k^2}\right] \, ,\quad m_R =
  \frac{N}{2}\frac{\m- \n}{k} \, .
\end{align}
Thus the mass above the ground state and the momentum, 
\begin{equation}
\label{eq:E-P}
\Delta E \,=\, \frac{\Delta h + \Delta \bar h}{R} \,=\, \frac{N}{R}\frac{\m^2 + \n^2 -1}{2k^2}  \, ,\qquad P \,=\, \frac{h-\bar h}{R} 
\,=\, \frac{N}{R}\frac{\m \n}{k^2} \;,
\end{equation}
match the ADM mass above the ground state~\eq{ADMmasslargeR} and momentum charge \eq{eq:Qpmn} derived from the supergravity solutions, after the standard conversion of units~\cite{Chakrabarty:2015foa}.

Note that there is a consistency condition in the symmetric product orbifold CFT that the momentum per strand be an integer~\cite{Dijkgraaf:1996xw}. This implies that $\m\n/k \in \mathbb{Z}$.


\section{Worldsheet CFT}
\label{sec:WS CFT}

In this section we review the construction of~\cite{Martinec:2017ztd}
which gives an exactly solvable
worldsheet description of the JMaRT solution, in the NS5 decoupling limit, in terms of a null-gauged
WZW model.

\subsection{Round supertubes}

As reviewed above, the JMaRT solution is obtained in the spacetime CFT by performing a
left/right asymmetric spectral flow from a round NS5-F1 supertube. It
is useful to perform a T-duality along the
$y$ direction, which brings us to the NS5-P duality frame.%
\footnote{Treatments in the literature often start from the D1-D5 duality frame, related to the NS5-F1 frame by S-duality.} 
We can now construct the NS5-P round supertube, as follows: We start from a configuration
of NS5 branes separated on their Coulomb branch, arranged
symmetrically on a circle, and then spin the branes up into a helical profile which
is kept from collapsing by adding momentum and angular momentum. The
$\mathbb{Z}_{n_5}$ symmetric
configuration of NS5 branes on a circle admits an exact treatment in
worldsheet string theory as an orbifold of a product of coset models~\cite{Sfetsos:1998xd}:
\begin{equation}\label{cosetorbifold}
\left(\frac{SL(2,\mathbb{R})_{n_5}}{U(1)} \times
\frac{SU(2)_{n_5}}{U(1)} \right) \Big/ \mathbb{Z}_{n_5} \, .
\end{equation}
An equivalent description, that makes clear the connection with the
target space solution, employs the gauged WZW model~\cite{Israel:2004ir}
\begin{equation}
\frac{SL(2,\mathbb{R})_{n_5} \times SU(2)_{n_5}}{U(1)_L \times U(1)_R}
\, ,
\end{equation}
where the gauged symmetry is generated by null currents in
$SL(2,\mathbb{R})\times SU(2)$:
\begin{equation}
U(1)_L : \quad \mathcal{J} = J_3^{sl} + J_3^{su} \, ,\quad U(1)_R :
\quad \bar{\mathcal{J}} = \bar J_3^{sl} + \bar J_3^{su} \, .
\end{equation}
 By using a parafermion decomposition
$SL(2,\mathbb{R}) = U(1)\!\times\! \frac{SL(2,\mathbb{R})}{U(1)}$, $SU(2) = U(1)
\!\times\! \frac{SU(2)}{U(1)}$, one can check that the gauging removes the scalars
parametrising the additional $U(1)$ directions, and one is left with
the product of the cosets in \eqref{cosetorbifold}, with a linear
constraint among their quantum numbers that corresponds to the action
of the $\mathbb{Z}_{n_5}$ orbifold.

\begin{figure}[ht]
\centering
    \includegraphics[width=0.6\textwidth]{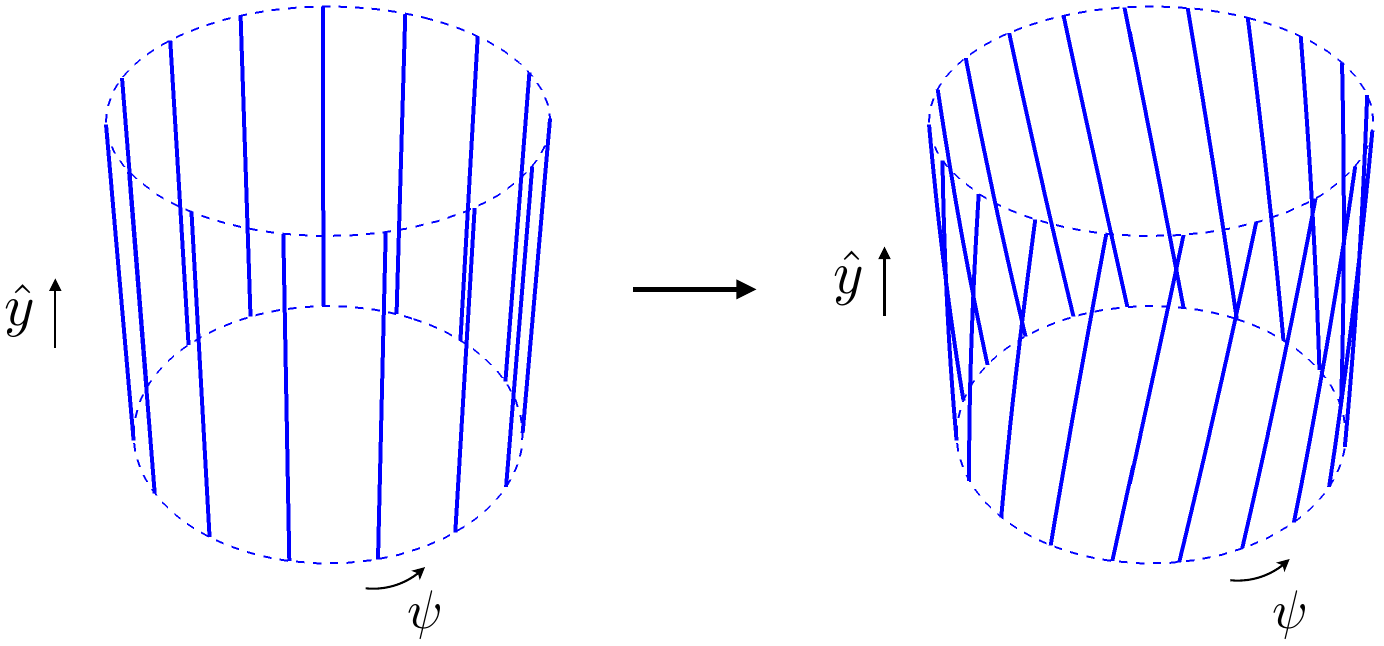}
\caption{\it 
On the left, a symmetric configuration of NS5 branes on their Coulomb
branch. Tilting the branes along the $(\ydual, \psi)$ torus, with a
monodromy that links the various strands together, and adding
momentum to keep the branes puffed up, gives an NS5-P supertube. In
worldsheet string theory, the tilting corresponds to a rotation of the
null current being gauged inside $SL(2,\mathbb{R})\times SU(2)\times \mathbb{R}^{1,1}$.
}
\label{fig:NS5Psupertube}
\end{figure}

In order to build the NS5-P
supertube, we consider a modification of the above construction, starting
with a WZW model on the group $\mathcal{G} =
SL(2,\mathbb{R}) \times SU(2) \times \mathbb{R}^{1,1}_{(t,\ydual)}$. We then gauge a
current that has a null component in $\mathbb{R}^{1,1}$ and we
compactify the~$\ydual$ direction. The target space of such a gauged WZW
model is the Lunin-Mathur solution for the round NS5-P supertube~\cite{Lunin:2001fv} (see Figure
\ref{fig:NS5Psupertube}).

Changing vector to axial gauging in the $\mathbb{S}^{1}_\ydual$ direction corresponds to a T-duality along $\ydual$ and the gauged WZW model corresponds to the round supertube in the NS5-F1 frame (with the T-dual coordinate $y$ parametrizing the circle).  The monodromy of the fivebrane source in the NS5-P frame translates to a monodromy of vanishing cycles of a KK-dipole loop in the NS5-F1 frame, such that as one goes once around the $y$ circle, one cyclically permutes the vanishing cycles.

Finally, as we will detail in the next section, the three-charge spectrally flowed supertubes are obtained in this
framework by considering a more general left/right asymmetric gauging,
where the currents $(\mathcal{J}, \bar{\mathcal{J}})$ are not
separately null in $SL(2,\mathbb{R})\times SU(2)$ and
in $\mathbb{R}^{1,1}$.

\subsection{Asymmetric null gauging and spacetime spectral flow}

In order to construct an exact model for strings propagating on the
JMaRT solution, we need to consider a supersymmetric WZW model with a target space
$\mathcal{G} \times \mathcal{M}$. We will take $\mathcal{M} = T^4$ and $\mathcal{G} =
SL(2,\mathbb{R}) \times SU(2) \times \mathbb{R}^{1,1}$, and we will focus
on the dynamics on $\mathcal{G}$. Note that the target space is a
twelve-dimensional group manifold with signature $(10,2)$. We denote
elements of $\mathcal{G}$ by $ (g_{\sl}, g_{\su}, \mathbf{x})$, where\footnote{We note that we have changed some conventions from those of~\cite{Martinec:2017ztd}, to be compatible with the conventions of~\cite{Jejjala:2005yu,Chakrabarty:2015foa}. We give a summary of the conventions used in this paper in Appendix \ref{app:conventions}.\label{foot:conventions}}
\begin{equation}\label{groupelements}
g_{\sl}= e^{\frac{i}{2}(\tau -\sigma) \sigma_3} e^{\rho \sigma_1}
e^{\frac{i}{2}(\tau+\sigma)\sigma_3} \, ,\quad g_{\su} =
e^{\frac{i}{2}(\psi -\phi) \sigma_3 }e^{i\theta  \sigma_1}
e^{\frac{i}{2}(\psi+\phi)\sigma_3} \, , \quad \mathbf{x} = (t,y) \, ,
\end{equation}
and where $\sigma_i$ $(i=1,2,3)$ are the Pauli matrices.\footnote{Explicitly, we take $\sigma_1 = \begin{pmatrix}  0 & 1\\ 1 & 0  \end{pmatrix}\,
  ,\quad  \sigma_2 = \begin{pmatrix}  0 & -i\\ i & 0  \end{pmatrix} \,
  ,\quad \sigma_3 = \begin{pmatrix}  1 & 0\\ 0 & -1  \end{pmatrix}$.} 
We work in conventions in which $\alpha'=1$. The WZW model on a group
manifold parametrized by $g$ is described by the action 
\begin{equation}
\mathcal{S}_{WZW}(g,\k)= \frac{\k}{2\pi}\int \text{Tr}
\left[(\partial g) g^{-1} (\bar{\partial} g) g^{-1} \right]+ \Gamma_{WZ}(g) \,.
\end{equation}
The action for our model is a sum $\mathcal{S}_{WZW}^{SL(2)\times SU(2)} =
\mathcal{S}(g_{\sl},n_5\tight+2) + \mathcal{S}(g_{\su},n_5\tight-2)$, supplemented by the
trivial action for $\mathbb{R}^{1,1}$, plus 8 free fermions to implement worldsheet supersymmetry. With the parametrization
\eqref{groupelements}, at leading order in the semiclassical large
$\nfive$ limit, the action reads:
\begin{align}
\label{Swzw}
\cS_{\sst \rm WZW}^{\mathcal{G}} &= \frac{\nfive}{\pi}\int d^2\zhat \Bigl[ 
D\rho \Dbar\rho + \sinh^2\! \rho \,D\sigma \Dbar\sigma - \cosh^2\!\rho
                                   \left( D\tau \Dbar\tau +
                                   D\tau\Dbar\sigma \tight - D\sigma\Dbar\tau\right)
\nn\\ & \hskip 2.5cm+
D\theta \Dbar\theta + \sin^2\!\theta \,D\phi \Dbar\phi  +
        \cos^2\!\theta\left( D\psi \Dbar\psi + D\phi\Dbar\psi \tight-D\psi\Dbar\phi\right) \Bigr]
  \nn \\
&\hskip 0.6cm +\frac{1}{\pi}\int d^2\hat z \left( -Dt \bar D t + Dy \bar D y \right) ,
\end{align}
where $d^2\zhat$ is the $\cN\tight=1$ superspace measure. The action $\mathcal{S}(g,\k)$ has a set of conserved left and right moving currents%
\footnote{At the classical level, at large $\nfive$.  For the precise definition of the currents and their operator algebra at the quantum level, including the fermion contributions, see Appendix~\ref{app:WZWmodels}.}
\begin{equation}
J_a= \k\, \text{Tr} \left[T_a (\partial g) g^{-1} \right] 
\, , \quad 
\bar J_a =  \k\, \text{Tr} \left[(T_a)^{\ast} g^{-1}\bar{\partial} g \right] \, .
\end{equation}
The prescription of~\cite{Martinec:2017ztd} to obtain the spectrally flowed supertube is to gauge a null $U(1)_L\times U(1)_R$ subgroup of the
affine $\mathcal{G}_L \times \mathcal{G}_R$ symmetry of the full
model, generated by the following combinations of left and right currents for the product factors:
\begin{align}
U(1)_L : \quad \mathcal{J} &= l_1 J^{\sl}_3 + l_2 J^{\su}_3  + l_3 \hat{P}_{t,L} +
  l_4 \hat{P}_{y,L} \, , \\
U(1)_R: \quad \bar{\mathcal{J}}& = r_1 \bar J_3^{\sl} + r_2 \bar  J_3^{\su}  + r_3 \hat{P}_{t,R} +
  r_4 \hat{P}_{y,R} \,,  \nonumber
\end{align}
where $\hat{P}_{t,L}\equiv-\partial t$, $\hat{P}_{t,R}\equiv-\bar\partial t$, $\hat{P}_{y,L}\equiv\partial y$, $\hat{P}_{y,R}\equiv\bar\partial y$. 
For suitable choices of left and right null vectors,
null gauging eliminates one timelike and one spacelike direction from
the target space, bringing us from a (10+2)-dimensional target space to
a (9+1)-dimensional physical subspace without closed timelike curves or
other pathologies.
The null conditions 
\begin{equation}
\label{eq:nullconstr}
0 = \langle \mathbf{l}, \mathbf{l} \rangle = n_5 (-l_1^2 +l_2^2)-l_3^2
+l_4^2  
~~,~~~~
0 = \langle \mathbf{r}, \mathbf{r} \rangle = n_5 (-r_1^2+r_2^2) -
r_3^2+r_4^2  
\end{equation}
ensure anomaly cancellation and independence of the left and right
gaugings. In our case we have explicitly
\begin{align}
\label{slsucurrents}
J^{\sl}_3 &= n_5(\cosh^2\!\rho \, D\tau + \sinh^2\!\rho \,D\sigma) \,  ,&\quad 
       J^{\su}_3 & =n_5( \cos^2\!\theta \, D\psi - \sin^2\!\theta \, D\phi) \,  ,     \nn\\
\bar J^{\sl}_3&= n_5(\cosh^2\! \rho \,\Dbar\tau - \sinh^2\!\rho \, \Dbar
                \sigma ) \, ,&\quad \bar J^{\su}_3 & = n_5 (\cos^2\! \theta \,\Dbar \psi +
            \sin^2\!\theta \, \Dbar \phi) \,,
\end{align}
and the action for the gauged model is
\begin{equation}\label{gaugemodelaction}
\cS_{gWZW}^{\mathcal{G}} \,=\, \mathcal{S}_{WZW}^{\mathcal{G}} +
\frac{1}{\pi}\int d^2\hat z \left[ \mathcal{A} \bar{\mathcal{J}}+\bar{\mathcal{A}}\mathcal{J}
  - \frac{\Sigma}{2} \bar\cA \cA \right] ,
\end{equation}
where
\begin{equation}\label{gaugedmodel}
\Sigma \,=\, n_5 (l_1 r_1 \cosh 2\rho - l_2 r_2 \cos 2\theta) +l_3 r_3 - l_4
r_4 \, .
\end{equation}

We choose a convenient parametrization of the coefficients in terms of
rapidities $(\zeta, \, \xi)$ and $(\bar{\zeta}, \bar{\xi})$ that ensures
the constraints \eqref{eq:nullconstr}:
\begin{align}
l_1 &= -{\mu} \sinh \zeta \, ,\quad l_2 =-\mu \cosh \zeta ~ ,
\quad l_3 = \sqrt{n_5}\,\mu \cosh \xi ~ ,\quad l_4 = - \sqrt{n_5}\,\mu \sinh\xi ~
  ,\nonumber \\
r_1 & =-{\mu} \sinh \bar \zeta \, ,\quad r_2 =-{\mu} \cosh \bar \zeta \, ,\quad 
r_3 =  \sqrt{n_5}\,\mu \cosh \bar \xi \, ,\quad r_4 = + \sqrt{n_5}\,{\mu}\sinh
      \bar \xi  ~. 
\end{align}
We can fix the gauge freedom in the action \eqref{gaugemodelaction} by
setting $\tau=\sigma=0$.
As shown in~\cite{Martinec:2017ztd}, after integrating out the gauge
fields, 
the target space geometry matches the NS5 decoupling limit of the JMaRT solution
\eqref{eq:CFTmetric}--\eqref{eq:CFT-dil}, with the following
dictionary between the JMaRT parameters and the boost parametrization:
\begin{equation}\label{eq:dictionary}
 \mu^2 = \frac{M}{2n_5} \, ,\quad \xi = \delta_1 - \delta_p \, , \quad \bar \xi =
  \delta_1 + \delta_p \, , \quad 
e^{2\zeta}  = \frac{\m+\n+1}{\m+\n-1} \, ,\quad  e^{2\bar \zeta} =
\frac{\m-\n+1}{\m-\n-1}  \, .
\end{equation}
From \eq{eq:charges} the charges $Q_1$ and $Q_p$ are
\begin{equation}\label{chargesxi}
Q_1 = \frac{M}{2}\sinh(\xi+\bar\xi) \, ,\qquad Q_p =
\frac{M}{2}\sinh(\bar\xi-\xi) \, .
\end{equation}
Since the parameters $\m$, $\n$ are related to the spectral flow
parameters as in \eqref{mnssbar},
the last two relations in \eqref{eq:dictionary} relate the rapidities $(\zeta, \bar \zeta)$ to $(s,\bar s)$. The dependence of $(\xi,
\bar \xi)$ on the spectral flow data can be found by using the
following relations:
\begin{align}\label{eq:sinhzeta}
\sinh \zeta  & = \left[ (\m+\n)^2 -1\right]^{-1/2}
=\frac{1}{\sqrt{M}}\frac{a_1-a_2}{\m+\n} = \frac{\sqrt{Q_5} Q_1}{M k R
               \cosh\bar\xi\hspace{-2mm}\phantom{\big\|}}  \, ,\cr
\sinh \bar{\zeta} & = \left[ (\m-\n)^2 -1\right]^{-1/2}
=\frac{1}{\sqrt{M}}\frac{a_1+a_2}{\m-\n} = \frac{\sqrt{Q_5} Q_1}{M k R
               \cosh\xi}\, , \\[.3cm]
\cosh \zeta  & = (\m+\n) \sinh \zeta \, ,\quad \cosh \bar{\zeta} = (\m-\n)
               \sinh \bar{\zeta}\nn \,.
\end{align}
From these, taking into account \eqref{chargesxi}, we obtain
\be
\sinh \xi \,=\, \frac{\hat \gamma^2 - \m \n}{\hat \gamma \sqrt{(\m+\n)^2
            -1}} \, ,  \label{eq:sinhxi}
						\qquad~
\sinh \bar\xi  \,=\, \frac{\hat \gamma^2 + \m \n}{\hat \gamma \sqrt{(\m-\n)^2
            -1}} \,,
\ee
where we use the shorthand
\begin{equation}\label{gammahat}
\hat \gamma \,\equiv\, \frac{ k R}{\sqrt{Q_5}} \, .
\end{equation}
%

\subsection{Physical state constraints}

The perturbative string spectrum of the model is determined by
standard techniques. In terms of the parafermion decomposition of
$SL(2,\mathbb{R})$ and $SU(2)$ operators, the gauging eliminates the
scalars corresponding to a combination of the $\mathcal{Y}_{sl}$, $\mathcal{Y}_{su}$
directions (see Appendix \ref{app:WZWmodels}) and external directions, imposing a linear constraint on
the corresponding quantum numbers~\cite{Martinec:2017ztd}:
\begin{align}
0 & \;=\;  l_1 (2\msl + n_5 \wsl) + l_2 (2\msu+n_5
    \wsu)  + l_3 E  +  l_4 P_{y,L} \, ,  \nn \\
0 & \;=\;  r_1  (2 \bmsl + n_5 \bwsl) + r_2 (2\bmsu+ n_5 \bwsu) + r_3 E  +  r_4 P_{y,R} \, . \label{jmartconstr}
\end{align}
Here $\msl$ and $\msu$ are the eigenvalue of $J_{sl}^3$ and
$J_{\su}^3$ respectively and $\wsl$, $\wsu$ are spectral
flow parameters (see Appendix \ref{app:WZWmodels} for a review); similarly for the right-moving quantum numbers.
We also define
\begin{equation}
P_{y,L/R} = \frac{n_y}{R} \pm  w_y R ~ .
\end{equation}
In the physical state spectrum, the role of the null constraints is to relate the energy and momentum along $t,y$ to that in $\sltwo$ along $\tau,\sigma$, with an admixture of angular momentum along $\sutwo$.

We record the following useful relations: 
\begin{align}
\label{ratios}
- \frac{l_3}{l_1} & \,=\,- \frac{r_3}{r_1} \,=\, \frac{M k R}{Q_1}
                    \cosh  \xi  \cosh
                  \bar\xi  \,=\, k R \varrho\, ,\nn\\
 \frac{l_4}{l_1} &\,=\, \frac{M k R}{Q_1} \sinh \xi \cosh \bar \xi   \,=\, kR (1-\vartheta) \,, \\
 {-} \frac{r_4}{r_1} & \,=\, \frac{M k R}{Q_1} 
                    \cosh  \xi \sinh \bar \xi \,=\, kR (1+\vartheta) \nn \, ,
\end{align}
where we introduced the parameters
\begin{equation}\label{varrhovartheta}
\varrho \,=\, \frac{M}{Q_1}(c_1^2c_p^2-s_1^2s_p^2)
\,=\,\frac{M}{Q_1}\cosh \xi
          \cosh \bar \xi \, , \qquad
\vartheta  \,=\, \frac{Q_p}{Q_1} \, .
\end{equation}
Using \eq{chargesxi}--\eqref{eq:sinhxi} we can express
these quantities in terms of $\m$, $\n$ and therefore in terms of the spectral
flow parameters $s$ and $\bar s$:
\begin{equation}
\varrho  = \frac{1}{\hat \gamma^2}\sqrt{(\m^2+\hat \gamma^2) (\n^2+\hat
  \gamma^2)-\hat \gamma^2} 
  ~~ ,~~~~ 
\vartheta  = \frac{\m \n}{\hat \gamma^2} ~ ,
\end{equation}
with $\hat \gamma$ given in \eqref{gammahat}.
We finally arrive at the following form for the null constraints:
\begin{align}
\label{jmartconstrclean}
-(2\msl+n_5 \wsl) - (\m+\n) (2\msu +n_5 \wsu) + k R \varrho E - k R (1-\vartheta)
  P_{y,L} &=0\, , \\
-(2\bmsl+n_5 \bwsl) - (\m-\n) (2\bmsu+n_5 \bwsu) + k R \varrho E + k R (1+\vartheta)
  P_{y,R} &= 0 \, . \nn 
\end{align}
The difference of these two constraints involves only integer quanta, provided that $\m\n\in k\bZ$; recall that this constraint is required in the spacetime CFT, as discussed at the end of Section~\ref{subsec:holographicdescription}.

For the supersymmetric backgrounds with $\m= s+1$, $\n =s$ we find
\begin{equation}
\varrho = 1 + \vartheta \,,~~~~~~  \vartheta =
\frac{s(s+1)}{\hat \gamma^2} \, ,
\end{equation}
and the constraints reduce to 
\begin{align}
-2\msl-n_5 \wsl - (2s+1) (2\msu +n_5 \wsu) + kR(E -P_{y,L}) + \frac{n_5 s(s+1)}{kR}(E+P_{y,L})&=0\, ,\nn\\
-2\bmsl-n_5 \bwsl - (2\bmsu+n_5 \bwsu) + kR(E + P_{y,R}) + \frac{n_5
  s(s+1)}{kR}(E+P_{y,L})&= 0 \, , \label{GLMTconstr}
\end{align}
which were obtained in
\cite{Martinec:2017ztd}.\footnote{In order to compare with Eq.\;(5.18)
  of \cite{Martinec:2017ztd} one must keep in mind the different conventions
  between JMaRT and the supersymmetric backgrounds studied in
  \cite{Giusto:2012yz}. See Appendix \ref{app:conventions} for details.}
Later we will
consider a large $R$ expansion, corresponding to a large and
approximately decoupled AdS throat as described in Section~\ref{sec:AdS}. The first terms in the large $R$ expansion for
$\varrho$ are
\begin{align}
\varrho &\;=\;1+ \frac{\m^2+\n^2-1}{2\hat \gamma^2} -
\frac{(\m^2+\n^2-1)^2 -4\m^2\n^2}{8\hat \gamma^4}+ \mathcal{O}(\hat
  \gamma^{-6}) \label{rhoexpansion}\\
&\;=\;1+\frac{s(s+1) + \bar s(\bar s+1)}{\hat \gamma^2} - \frac{2
  s(s+1)\bar s(\bar s+1)}{\hat \gamma^4} + \mathcal{O}(\hat
  \gamma^{-6}) \, . \nn 
\end{align}

In addition to the linear constraints from the null gauging, in order
to determine the spectrum one must impose the usual
Virasoro constraints for the model on $\mathcal{G}$, for instance in the NS sector: 
\begin{align}
\label{VirConstraints}
0 \,=\, L_0 - \frac12 \,&=\,  -\frac{\jsl(\jsl-1)}{n_5} +
\frac{\jsu(\jsu+1)}{n_5}  - \msl \wsl -
\frac{n_5}{4}\wsl^2	\nn\\
& \quad~ + \msu \wsu + \frac{n_5}{4}\wsu^2
-\frac14 E^2 +\frac14 P_{y,L}^2 + \NL \;,  \\[.3cm]
0 \,=\,\bar L_0 - \frac12 \,&=\,  -\frac{\jsl(\jsl-1)}{n_5} +
\frac{\jsu(\jsu+1)}{n_5}  - \bmsl \bwsl -
\frac{n_5}{4}\bwsl^2 \nn\\
&  \quad~ + \bmsu \bwsu+ \frac{n_5}{4}\bwsu^2
-\frac14 E^2 +\frac14 P_{y,R}^2 + \NR \;,  \nn
\end{align}
where $\hat N_{L,R} = N_{L,R}-\frac12$.

A large AdS$_3$ region in the effective geometry of the gauged WZW model arises at large $\Ry$.  Because the $\sltwo$ currents~\eqref{slsucurrents} are exponential in the radial coordinate $\rho$,  the crossover between the linear dilaton region occurs at a radial location $e^{2\rho}\propto \Ry$.  The role of the constraints~\eqref{jmartconstrclean} in the string spectrum is to relate the ``AdS$_3$ energy/momentum/winding'' $\msl\pm\bmsl$, $\wsl\pm\bwsl$, measured locally in the cap of the geometry, to the asymptotic energy/momentum/winding $E,P_{y,L},P_{y,R}$.

\section{A compendium of string states}
\label{sec:Spectrum}

In this section we lay out the spectrum of closed string states.  There are several features of interest:
\begin{itemize}
\item
The $\sltwo$ zero-mode spectrum consists of discrete series
representations $\mathcal{D}_j^\pm$ corresponding to bound states
localized near the cap of the geometry, and continuous series
representations $\mathcal{C}_j$ corresponding to scattering states.
Unitarity imposes the bound $\frac12\le \jsl \le \frac12(n_5+1)$ for discrete series representations.
\item
Additional states are obtained by $\sltwo$ spectral flow transformations, leading to string states winding around the $\sigma$ circle.  Only axial spectral flow transformations are allowed, because the target space of the $\sltwo$ WZW model is the universal cover of the group manifold, \ie\ both timelike directions of the target space are non-compact.
\item
The spectrum of unitary highest-weight representations of the $\sutwo$
current algebra consists of zero-mode representations of spin $0\le
\jsu \le \frac12 (n_5-2)$.  Spectral flow leads to additional states outside this bound.  In some sense these states have ``winding'' in the $\sutwo$ angles $\phi$ and $\psi$ that plays a role in the $\ads3\times\bS^3$ version of the giant graviton phenomenon~\cite{McGreevy:2000cw,Lunin:2002bj}.
\item
Large gauge transformations combine particular spectral flows in $\sltwo$ and $\sutwo$ with shifts of the zero-modes of $y$ and $t$, leading to equivalence relations among perturbative string states.  These equivalences relate winding around the $\sltwo$ angular direction $\sigma$ and winding around the $y$ circle.
\item
Winding around the $y$ circle carries the background F1 charge; scattering states with $w_y\ne 0$ describe background charge entering or leaving the system.
\item
The zero-mode eigenvalues entering the null constraints~\eqref{jmartconstrclean} are all real; therefore there can be no instability of the background at the linearized level in the NS5 decoupling limit, in contrast to the linearized instability seen in the full asymptotically flat JMaRT solutions~\cite{Cardoso:2005gj,Chowdhury:2007jx}. This is consistent with the fact that in the NS5 decoupling limit there exists a globally timelike Killing vector field, as shown in Appendix \ref{app:ergoregion}.
\end{itemize}
\noindent

\subsection{Supergravity}
\label{sec:supergravity}

We begin with states in the supergravity spectrum.
In the worldsheet (NS,NS) sector, supergravity modes have $\NL=\NR=0$ by virtue of having a fermion of each chirality to carry  a tensor polarization. The Virasoro highest-weight constraints can be solved by setting these polarizations to lie along $\cM$, yielding a set of minimally coupled scalars in six dimensions.  For simplicity, we focus on these modes; more generally the constraints tie the polarization state to the quantum numbers of the six-dimensional zero modes. In more detail, if one takes the ten-dimensional string-frame metric \eq{metricJMaRT} and simply discards the internal directions along along $\cM$, one obtains the six-dimensional Einstein-frame metric. Then, for example, an appropriately normalized metric fluctuation along $\cM$ reduces to a minimally coupled scalar on the background of the six-dimensional Einstein metric (see e.g.~\cite{David:2002wn}).

We begin with the states localized at the
cap of the geometry. These states descend from the discrete series
representations $\mathcal{D}_j^\pm$ of $SL(2,\mathbb{R})$ which have been used to describe bound states in the cigar coset $SL(2,\mathbb{R})/U(1)$~\cite{Giveon:1999px,Giveon:1999tq}. In the supergravity
limit, the spectrum obtained from the worldsheet model agrees with the
spectrum of the frequencies obtained by solving the wave equation in the
JMaRT geometry.

\subsubsection{Bound states}
\label{subsec:SugraBoundStates}

We start with all winding quantum numbers turned off, since for large $n_5$ we know that these states will match low-lying supergravity states when $\jsl,\jsu\ll n_5$.  These lowest-energy modes correspond to supergravity modes at the bottom of the cap in the geometry. 
The principal discrete series representations of
$SL(2,\mathbb{R})$, $\mathcal{D}_{j}^+$, and their conjugates,
$\mathcal{D}_j^-$, describe states with power-law decay at large $r$. The corresponding string states
are bound states that live close to the tip of the cigar; in the full geometry, these states live near the cap. The energy spectrum is determined
directly from the gauge constraints \eqref{jmartconstrclean}. We
initially consider the sector with vanishing spectral flow quantum numbers $\wsl=\wsu=\bwsl=\bwsu=0$, and zero winding along the $y$ circle, $w_y = 0$, so that $P_{y,L} = P_{y,R} =
n_y/R$. 
We also define the quantities 
\begin{align}
\mathcal{S} &= E \varrho k R + n_y k\vartheta - \m (\msu + \bmsu) -
              \n (\msu - \bmsu) \, , \label{eq:defSD}\\
 \mathcal{D} & = k n_y + \m (\msu - \bmsu) +
              \n (\msu + \bmsu)  \, . \nonumber
\end{align} 
The constraints \eqref{jmartconstrclean} can then be written as
\begin{equation}
\label{nullconstrs}
2\msl - \mathcal{S} + \mathcal{D} = 0 \, ,\quad 2\bmsl -\mathcal{S} -
\mathcal{D} = 0 \, .
\end{equation}
In the unflowed sector, the condition $w_y=0$ ensures that the $L_0 -
\bar L_0$ constraint is satisfied. The $L_0 + \bar L_0$ constraint
gives the condition for massless states
\begin{equation}
\label{sugraMassShell}
-\frac{\jsl(\jsl-1)}{n_5} +\frac{\jsu(\jsu+1)}{n_5} - \frac{E^2}{4} +
\frac{n_y^2}{4 R^2} \,=\, 0  \, ,
\end{equation}
which fixes the $SL(2,\mathbb{R})$ spin to be
\begin{equation}
\label{eq:jslsugra}
\jsl \,=\, \frac{1}{2}+\frac{1}{2}\sqrt{1+\Lambda - n_5 \left( E^2 -
    \frac{n_y^2}{R^2}\right)} \,\equiv\, \frac{1}{2} + \frac{\nu}{2}\, ,
\end{equation}
where we defined
\begin{equation}
\Lambda \,=\, 4 \jsu (\jsu+1) \, .
\end{equation}
If we take the unitary discrete series representation
$\mathcal{D}_j^{+}$
we have 
\begin{equation}
\msl =  \jsl + n \, , \quad~~  \bmsl =  \jsl +\bar n \, ,\qquad n\, , \bar n \in \mathbb{N} \,.
\end{equation}
One may then solve \eg\ the first of the null gauge constraints~\eqref{nullconstrs} with
\begin{equation}\label{eq:nullconstSD}
\nu+1 - \mathcal{S}  + \mathcal{D} = -2 n \, , \qquad n \in
\mathbb{N} \, ,
\end{equation}
and use the second constraints to fix $\bar n$ from
\begin{equation}
\mathcal{D} = \bar n - n \, .
\end{equation}
The energy of these states is given by
\begin{equation}
k R \varrho E = \nu+1 + n_y k (1-\vartheta)  +2 \msu (\m+\n) + 2n\, .
\end{equation}
It is interesting to compute the spectrum in the  large $R$ limit, in
which the solution has a large $\ads3\times \bS^3$ region. From
\eqref{rhoexpansion} we see that the leading solution has $\varrho
= 1 + \cO(R^{-2}) $, $\vartheta = \cO(R^{-2})$. Hence we have
\begin{equation}\label{D+sugraconstr}
k R E = 2 \jsl+ n_y k  + 2 \msu (\m+\n) +2n\, ,\qquad \jsl = \jsu + 1
\, .
\end{equation}

Alternatively, we can consider the $\mathcal{D}_j^{-}$ representation, for which  
\begin{equation}
\msl =  -\jsl - n \, , \quad~~  \bmsl = - \jsl -\bar n \, ,\qquad n\, , \bar n \in \mathbb{N}\,.
\end{equation}
We choose to solve the second null constraints~\eqref{nullconstrs} for
$\mathcal{S}$ to find
\begin{equation}
\label{nbareq}
\nu + 1 + \mathcal{S} + \mathcal{D} = - 2\bar n\, , \quad \bar n \in
\mathbb{N} \, ,
\end{equation}
with the first constraint giving
\begin{equation} \label{eq:D-Dplus}
\mathcal{D} = n - \bar n \, .
\end{equation}

An interesting feature is revealed by this (axial) null constraint \eq{eq:D-Dplus}, which we can write as
\be 
\label{SugraAxialNullConstraint}
n-\bar n = k n_y 
+(\m+\n)\msu
-(\m-\n)\bmsu  ~.
\ee
Given a solution for a particular $y$ momentum $n_y$ and angular momenta $\msu,\,\bmsu$, a change in the asymptotic $y$ momentum by one unit $\delta n_y =1$ must be compensated by a change in the momentum on the $\sigma$ circle by $k$ units, $\delta(n\tight-\bar n)=k$.  This is rather suggestive of the physics of momentum fractionation observed in the symmetric product orbifold, which is the spacetime CFT in a weak-coupling region of its moduli space.  There, supergravity modes have integer momenta, while the natural moding of excitations along the strands of length $k$ is in units of $1/k$.

\subsubsection{Scattering states}

The AdS regime corresponds to energies $E \sim O(\Ry^{-1})$, a regime where states are localized in the cap region; at energies of order $\Ry^0$, one begins to encounter perturbative string states that are not bound to the cap, but rather have plane wave asymptotics at large $r$.  Such plane waves are described by continuous series representations $\cC_j$, where $\nu$ in~\eqref{eq:jslsugra} is pure imaginary.  In~\eqref{eq:jslsugra}, one thus has the energy $E$ dominating the remaining terms under the square root,
\be
\label{Elowerbound}
E>\sqrt{\frac{\bigl(2\jsu\!+\!1\bigr)^2}{\nfive}+ \frac{n_y^2}{\Ry^2}} ~,
\ee
where we note in particular that $E>1/\sqrt\nfive$.  These $\sltwo$ representations have
\begin{equation}
\msl =  \alpha + n 
~ , ~~~  
\bmsl =  \alpha +\bar n 
~;~~~
n\, , \bar n \in \mathbb{Z}
~,~~
0\le\alpha<1  ~.
\end{equation}
The null gauge constraints \eqref{nullconstrs} are then
\begin{align}
2(n+\alpha)-\cS+\cD &= 0 \nn\\
2(\bar n+\alpha)-\cS-\cD &= 0 ~.
\end{align}
Note that because $E$ is of order $\Ry^0$, $\cS$ and therefore $n+\bar n$ are $O(\Ry)$; in particular, the energy scale $n+\bar n$ in the $\sltwo$ timelike direction is of order the UV cutoff of the AdS region (or larger).

\subsubsection{Scalar wave function}

The $L_0+\bar L_0$ constraint is the string version of the wave equation.  This motivates comparison of the above results with the solutions to the wave equation for a 6D minimally coupled scalar in the JMaRT solution~\cite{Cvetic:1997uw,Jejjala:2005yu,Cardoso:2005gj,Chakrabarty:2015foa}. We now review this relation, returning temporarily to the full asymptotically flat solutions~\eq{metricJMaRT}. The wave equation is:
\begin{equation}\label{scalarwavefunction}
\Box \Psi = \frac{1}{\sqrt{-g}} \partial_{\mu} \Big( \sqrt{-g} g^{\mu
  \nu} \partial_{\nu} \Psi \Big ) = 0 \, .
\end{equation}
Starting with the full solution without taking the fivebrane decoupling limit, one can separate variables by taking the ansatz
\begin{equation} \label{scalarwavefunction-2}
\Psi = \exp \Big[ - i \omega t + i m_{\psi} \psi + i m_{\phi} \phi + i
\frac{\lambda}{R}y\Big] \chi(\theta) h(r) \, .
\end{equation}
The angular and radial wave equations are then
\begin{align}
\frac{1}{\sin 2\theta} \frac{d}{d\theta} \Big(\sin 2\theta
  \frac{d}{d\theta}\chi\Big) + \Big[ \Big( \omega^2 -
  \frac{\lambda^2}{R^2}\Big) (a_1^2\sin^2\theta + a_2^2 \cos^2\theta )
  - \frac{m_{\psi}^2}{\cos^2\theta} - \frac{m_{\phi}^2}{\sin^2\theta}
  \Big] \chi & = -\Lambda \chi \, , \nonumber \\
4 \frac{d}{d x}\Big[x  \Big(x + k^{-2})\frac{d h}{d x} \Big] + \Big[
  \kappa^2 x + 1 -\nu^2+\frac{\sxi^2}{k^2 x + 1} -
  \frac{\dzeta^2}{k^2x}\Big] h & = 0 \, , \label{waveequation}
\end{align}
where 
\begin{equation}
x = k^{-2}\sinh^2 \rho \, .
\end{equation}
The parameters are defined as\footnote{Our notation is related to that of \cite{Chakrabarty:2015foa} by $\mathcal{S} = \xi k$, $\mathcal{D} = \zeta k$.}
\begin{align}
\kappa^2 &= \Big(\omega^2 - \frac{\lambda^2}{R^2} \Big) (r_{+}^2-r_{-}^2)k^2
  \, , \label{wavefunctionparameters} \\
\sxi &= \omega \varrho k R - \lambda k\vartheta - m_{\phi} \n + m_{\psi} \m
      \, , \nonumber\\
\dzeta & =  - k \lambda - m_{\psi} \n + m_{\phi} \m \, , \nonumber\\
\nu^2 & = 1 + \Lambda - \Big( \omega^2 -\frac{\lambda^2}{R^2}\Big) (r_{+}^2
        + M s_1^2 + M s_5^2) - \Big(\omega c_p - \frac{\lambda}{R}
        s_p\Big)^2 M \, ,
\end{align}
with
\begin{equation}
\varrho  \,=\, \frac{M^2}{Q_1 Q_5}\Big(c_1^2 c_5^2 c_p^2 - s_1^2
          s_5^2s_p^2\Big) \, ,\qquad
\vartheta \,=\, \frac{M Q_p}{Q_1 Q_5} \Big(c_1^2 c_5^2-s_1^2s_5^2\Big)  \,.
\end{equation}
In the NS5 decoupling limit, the $\cS$, $\cD$ defined in \eq{wavefunctionparameters} become equal to the corresponding quantities with the same names defined in
\eq{eq:defSD}, with the following map between parameters:  
\begin{equation} \label{eq:convention-map}
m_{\psi} =  -(\msu + \bmsu) \, ,\qquad m_{\phi} = (\msu - \bmsu) \, , \qquad n_y = - \lambda ~.
\end{equation}

Equation~\eqref{waveequation} can be solved perturbatively via matched asymptotic expansion~\cite{Cvetic:1997uw,Jejjala:2005yu,Cardoso:2005gj,Chakrabarty:2015foa}. In the NS5 decoupling limit described in Section \ref{sec:fivebranedecouplinglimit}, the equations simplify considerably. In the angular equation, assuming that $\omega \sim 1/R$ or $\omega \sim 1/\sqrt{Q_5}$, the terms proportional to $\omega^2$ and $\lambda^2$ drop out, and hence we find to leading order
\begin{equation}
\Lambda = l ( l+2) \, .
\end{equation} 
Under the same assumptions, in this limit one can also neglect the $\kappa^2 x$ term in the radial wave
equation, so that the equation reduces to
\begin{equation}\label{eq:6dscalar}
4 \frac{d}{d x}\Big[x  \Big(x + k^{-2})\frac{d h}{d x} \Big] + \Big[
1 -\nu^2+\frac{\sxi^2}{k^2 x + 1} -
  \frac{\dzeta^2}{k^2 x}\Big] h = 0 \, ,
\end{equation}
where to leading order the parameter $\nu$ is given by
\begin{equation}
\nu^2 = 1 + \Lambda - Q_5 \Big( \omega^2 -\frac{\lambda^2}{R^2}\Big) \,,
\end{equation}
while the parameters $\sxi$, $\dzeta$ are given by
\eqref{wavefunctionparameters} with 
\begin{align}
\varrho  = \frac{M}{Q_1}\Big(c_1^2c_p^2 - s_1^2
         s_p^2\Big) \, ,\qquad
\vartheta  = \frac{Q_p}{Q_1} \, . \nonumber
\end{align}
Note that these quantities $\varrho$, $\vartheta$ are the same as those introduced above
in \eqref{varrhovartheta}. 
The parameter $\nu$ also matches the definition \eqref{eq:jslsugra}
after taking into account the relation $l = 2 \jsu $.
Equation \eqref{eq:6dscalar} is hypergeometric, of the same form as the wave equation in
AdS. The solution that is regular at the cap ($x=0$) is given by
\begin{equation} \label{eq:h-sol}
 h (x) \,=\, C k^{-\sxi} x^{\frac{|\dzeta|}{2}} (1+k^2 x)^{\frac{\sxi}{2}} \,
 {}_2F_1\left(s_{-}, s_{+}
   , 1+|\dzeta| , -k^2x\right) \, ,
\end{equation}
where $C$ is a normalization coefficient and where
\begin{equation}
s_{\pm} = \frac12\left(1\pm\nu+ \sxi+|\dzeta|\right)  \, .
\end{equation}
Here again we see a reflection of mode fractionation in the cap along the lines mentioned at the end of Section~\ref{subsec:SugraBoundStates}~-- a change in the momentum moding $\delta n_y=1$ results in a shift $\delta s_\pm =\delta |\dzeta| =k$, and puts $k$ more nodes in the radial wavefunction.  It seems that the natural moding for waves in the cap, which would change the number of nodes by one, is not available to supergravity, consistent with the observations in~\cite{Bena:2016agb}.

At large $x$, the radial wavefunction behaves as
\begin{align}
\label{wavefnasymp}
h(x) &\,\sim\, \frac{ C k^{-1-\nu-\sxi-|\dzeta|}\Gamma(1+ |\dzeta|)
  \Gamma(-\nu)}{\Gamma\left(\frac12\left(1-\nu - \sxi+|\dzeta|\right) \right)
    \Gamma\left(\frac12\left(1-\nu + \sxi+|\dzeta|\right)  \right)} x^{\frac{-\nu-1}{2}} \nn\\
&\quad~~+\frac{C k^{-1-\nu- \sxi-|\dzeta|}\Gamma(1+ |\dzeta|)
  \Gamma(\nu)}{\Gamma\left(\frac12\left(1+\nu - \sxi + |\dzeta|\right) \right)
    \Gamma\left(\frac12\left(1 + \nu  + \sxi+|\dzeta|\right)  \right)}
  x^{\frac{\nu-1}{2}} \, .
\end{align}
To select perturbations that are bound to the tip we set to zero the
term that diverges at large $x$. This amounts to selecting a pole for one of the gamma
functions in the denominator, giving the two possibilities:
\begin{equation}\label{sugrapoleconditions}
1 + \nu  - \sxi+|\dzeta| 
= - 2 \hat{n}  \, , \quad~~ 1 + \nu  + \sxi+|\dzeta| = - 2\tilde{n} \, , \quad~~ \hat{n}, \tilde{n} \in \mathbb{N} \, .
\end{equation}
Let us compare these two conditions to the spectrum of excitations in the $\mathcal{D}_j^+$ and $\mathcal{D}_j^-$
representations. For the $\mathcal{D}_j^+$ representation, we have $\cD = \bar{n}-n$. Following the discussions in~\cite{Chowdhury:2007jx,Chakrabarty:2015foa}, if $\cD>0$ one can write~\eqref{eq:nullconstSD} as
\begin{equation}\label{eq:nullconstSD-2}
1+\nu - \mathcal{S}  + |\mathcal{D}| = -2 n \, , \quad n \in
\mathbb{N} \, ,
\end{equation}
and if 
$\cD<0$ 
one can write~\eqref{eq:nullconstSD} as
\begin{equation}\label{eq:nullconstSD-3}
1+\nu - \mathcal{S}  + |\mathcal{D}| = -2 \bar{n} \, , \quad \bar{n} \in
\mathbb{N} \, ,
\end{equation}
both of which match the first choice in~\eq{sugrapoleconditions}. Similarly, the $\mathcal{D}_j^-$ representation spectrum \eq{nbareq}, \eq{eq:D-Dplus} matches the second choice in~\eq{sugrapoleconditions}. Thus we find exact agreement.

Let us compare the local energy of a static observer in the $\ads3\times \bS^3$ cap to the energy $E$. Cap energy
is found by expanding the wave function $\Psi$ in \eqref{scalarwavefunction} in terms of coordinates in which
six-dimensional metric takes the standard $\ads3\times \bS^3$ form. Specifically, we separate variables similarly to \eq{scalarwavefunction-2} but in terms of the spectrally flowed angular coordinates $\tilde\psi$, $\tilde\phi$ defined in \eqref{AdS3variables}, as done in~\cite{Giusto:2012yz}. This gives cap energy and
momentum $\tilde \omega$, $\tilde \lambda$ as follows:
\begin{equation}
k R\tilde \omega = k R\omega + \m m_{\psi} - \n m_{\phi} \, , \quad
k \tilde \lambda = k \lambda + \n m_{\psi} - \m m_{\phi} \, .
\end{equation}
From \eqref{wavefunctionparameters}, we see that in the large $R$
limit we have $\cS = k R \tilde \omega$ and $\cD = - k \tilde \lambda$. This is precisely what we expect, since the null
constraints \eqref{nullconstrs} set
\begin{equation}
\msl + \bmsl = k R \tilde \omega \, ,\qquad \msl - \bmsl = k \tilde
\lambda \,,
\end{equation}
and so in particular the cap energy $\tilde\omega$ is proportional to the $\sltwo$ energy $\msl + \bmsl$.
For the $\mathcal{D}^-_j$ modes, we have $\msl = -(\jsl+n)$, $\bmsl = -(\jsl+\bar n)$, so the cap energy is negative, and similarly the $\mathcal{D}^+_j$ modes have positive cap energy.

Note that for supersymmetric backgrounds, the left- or right-moving energy on the same side as the supersymmetry always has the same sign for asymptotic and cap observers.
In terms of the left/right quantities \eqref{eq:convention-map} and the left/right
spectral flows the constraints for $\cD^-$ take the form
\begin{align}
\label{Dminusconstraints}
2n &= - k R \omega  - k\lambda + 2(2s+1) \msu - 2\jsu -2\, , \nn\\
2\bar n   &= - k R\omega + k  \lambda +2 (2\bar s +1) \bmsu -2\jsu -2\, .
\end{align}
where we set $\jsl = \jsu + 1$ (see \eqref{D+sugraconstr}). Let us
consider the supersymmetric backgrounds in which $\bar s = 0$. Recalling that
$|\msu|, |\bmsu| \leq \jsu$ and $n, \bar n \geq 0$, we find that the
right moving energy  $\omega +n_y/\Ry$ (recall $n_y=-\lambda$) is never
positive, and so there are no $\cD^-$ modes with positive asymptotic chiral energy.

Away from the decoupling limit, the full asymptotically flat, non-supersymmetric JMaRT solutions have the property that any choice of Killing vector field that is asymptotically causal (i.e.~timelike or null) becomes spacelike in the interior of the geometry. Smooth horizonless solutions with this property generically have an ergoregion instability~\cite{Friedman:1978}, provided that the ergoregion does not extend to infinity in any direction, as is the case for the asymptotically flat JMaRT solutions.  The instability arises from zero-norm modes that have their principal support in two places: A component in the ergoregion which has negative energy as measured from infinity, and a component in the asymptotically flat region that has positive energy as measured from infinity. This is related to the Schiff-Snyder-Weinberg effect~\cite{Schiff:1940}; for further discussion see~\cite{Chowdhury:2008bd} and references therein.

Supersymmetric smooth horizonless solutions are expected to be linearly stable; this class includes the supersymmetric limit of the JMaRT solutions~\cite{Giusto:2004id,Giusto:2004ip,Giusto:2012yz}, which have a globally null Killing vector field. In the analysis described above around Eq.\;\eqref{Dminusconstraints}, the absence of $\cD^-$ modes with positive asymptotic chiral energy is a consequence of this property of the supersymmetric backgrounds.

By employing a matched asymptotic expansion, one can show that the above $\mathcal{D}_j^{\pm}$ modes in the fivebrane throat (with power-law fall-off) match onto spherical Bessel functions (with plane-wave asymptotics) in the asymptotically flat region~\cite{Cvetic:1997uw,Jejjala:2005yu,Cardoso:2005gj,Chakrabarty:2015foa}. This analysis reveals that the allowed completions of the $\mathcal{D}_j^-$ modes are the unstable modes of the ergoregion instability -- the imaginary part of the frequency is positive, so the modes grow exponentially~\cite{Cardoso:2005gj}. This is consistent with our conventions that the $\mathcal{D}_j^-$ modes have negative energy in the cap, as described above. We note that the opposite sign eigenvalue of the local timelike Killing vector ($\tau$ in the cap; $t$ asymptotically) leads to an overall null Klein-Gordon norm for these modes. 

The resulting ergoregion emission has been interpreted holographically as a classically enhanced version of Hawking radiation from atypical microstates of the system~\cite{Chowdhury:2007jx,Chakrabarty:2015foa}, as follows: In the weak-coupling CFT, the matrix elements that describe the ergoregion emission and those that describe Hawking radiation from thermal states are of precisely the same sort. These matrix elements involve exactly the same vertex operators that couple the CFT state to radiating supergravitons~\cite{Avery:2009tu}; the only difference is that for the JMaRT solution the dual CFT state is an atypical coherent state, rather than a typical pure state of a thermal ensemble. 

In contrast, in the NS5 decoupling limit of the JMaRT solutions, there is always a globally timelike Killing vector field, as we show in Appendix \ref{app:ergoregion}. Therefore there is no ergoregion instability in the NS5 decoupling limit, and so we should not see the instability in the worldsheet CFT. We find that this is indeed the case.

\subsection{Worldsheet spectral flow}

Spectral flow in the worldsheet $\sltwo$ and $\sutwo$ WZW models generates additional states, starting from the current algebra highest weight states.  The worldsheet quantum numbers shift similarly to~\eqref{STspecflow}; for instance in $\sutwo$ at level $\nfive$ one has
\be
\label{WSspecflow}
\bigl(\delta J_\su,\delta \bar J_\su\bigr) =\frac \nfive2\,\bigl(\wsu,\bwsu\bigr)
\,,~~~~
(\delta L_0,\delta\bar L_0) =
\frac{\nfive}{4}\bigl(\wsu^2,\bwsu^2\bigr) + \bigl(\wsu  J_\su, \bwsu \bar J_\su\bigr)\,.
\ee

For $\sutwo$, spectral flow maps current algebra highest weight states of spin $\jsu$ to states which are typically not current algebra highest weight,%
\footnote{The exception being the vacuum, which maps to the current algebra highest weight state with $\jsu=n_5/2$ under spectral flow by $\wsu=1$.}
but which are highest weight under the zero-mode algebra with $\jsu>n_5/2$.  Since the physical state conditions for strings involve the conformal representation theory rather than the current algebra representation theory, such flowed states are perfectly acceptable as building blocks of physical states; the upshot is that the $\sutwo$ spin extends from $0\le \jsu\le \nfive/2$ to all values of $\jsu$. Indeed, the conformal dimension of such states is the naive extension of the highest weight expression
\be
L_0 = \frac{\jsu(\jsu+1)}{\nfive} +  \wsu \jsu + \frac{\nfive \wsu^2}4 = \frac{(\jsu+\nfive \wsu/2)(\jsu+\nfive \wsu/2+1)}{\nfive}-\frac{\wsu}{2}
\ee
up to a linear shift $\wsu/2$.

The states with nonzero $\wsu,\bwsu$ are related to the phenomenon of giant gravitons on the sphere%
~\cite{McGreevy:2000cw,Lunin:2002bj}.  
Classical solutions to the $\sutwo$ WZW model are
\be
g_\su =  g_l(z)g_r(\bar z)  ~;
\ee
pointlike strings traveling great circles on $\bS^3$ are given by 
\be
g_l(z) = h_l\, e^{iz\,\hat {\bf n}\cdot\boldsymbol{\sigma}/2}
\,,~~~~ 
g_r(\bar z) = e^{i\bar z\,\hat {\bf n}\cdot\boldsymbol{\sigma}/2} \, h_r 
\ee
where $h_l,h_r$ are constant $\sutwo$ matrices.  Spectral flow amounts to the transformation
(see \eg~\cite{Maldacena:2000hw})
\be
\label{ClassicalSpecFlow}
g_l\to e^{i\wsu z\,\sigma_3/2} g_l
\,,~~~~
g_r\to g_r \, e^{i\bwsu\bar z\,\sigma_3/2}  ~;
\ee
the $\phi$ and/or $\psi$ directions now have a nontrivial spatial dependence on the string worldsheet, 
and the string has ``puffed up'' along the sphere in a direction transverse to its angular momentum.

Spectral flow in $\sltwo$ is somewhat more intricate~\cite{Maldacena:2000hw,Maldacena:2001km,Giveon:2016dxe} (see~\cite{McElgin:2015eho} for a recent review).  The features relevant for us are quite similar to $\sutwo$ spectral flow. 
The semiclassical transformation~\eqref{ClassicalSpecFlow} introduces winding of the string worldsheet around the $\sigma$ and/or $\tau$ directions of the $\sltwo$ group manifold, and shifts the quantum numbers according to
\be
\label{SL2specflow}
\bigl(\delta J_\sl,\delta \bar J_\sl\bigr) =\frac
\nfive2\,\bigl(\wsl,\bwsl \bigr)
~,~~~~
(\delta L_0,\delta\bar L_0) = -\frac{\nfive}{4}\bigl(\wsl^2,\bar \wsl^2\bigr) - \bigl(\wsl J_\sl, \bwsl \bar J_\sl\bigr)~.
\ee
In this way, $\sltwo$ spectral flow sectors appear to describe the AdS$_3$ version of giant gravitons.  States in the continuous series of $\sltwo$ describe strings propagating out to infinity, and the spectral flow sectors are thus related to strings winding the AdS angular direction which are entering or leaving the cap.

As usual with giant graviton states, there is no topological spatial circle in either $\sutwo$ or $\sltwo$ (the brane tension is compensated by a combination of angular momentum and interaction with flux), and as a result the ``winding'' quantum numbers $\wsl,\bwsl,\wsu,\bwsu$ are not conserved in correlation functions~\cite{Maldacena:2001km,Giribet:2007wp}.

With $\sutwo$ spectral flow turned on,
the axial Virasoro constraint~\eqref{VirConstraints},
\be
L_0-\bar L_0 = \frac{\nfive}{4}\bigl( \wsu^2-\bwsu^2\bigr) +\bigl( \msu\wsu-\bmsu\bwsu\bigr) + \NL-\NR=0 \,,
\ee
is satisfied by $\wsu=\pm\bwsu$, $\msu=\pm\bmsu$, $\NL=\NR$, corresponding to round stretched strings winding along the $\phi$ direction while orbiting in the $\psi$ direction, or vice versa.
From~\eqref{ClassicalSpecFlow}, $\what$ units of spectral flow correspond to a worldsheet that is spiraling around like a supertube -- because it {\it is} a supertube, traveling the same path in spacetime $\what$ times, and supported by angular momentum, while it can also be thought of as having the $\what^{\rm th}$ mode excited in some plane.  In the usual  discussion of supertubes in flat spacetime, one matches winding/momentum along one chirality of the string with oscillators on the other side for level matching. By contrast, here the winding/momentum is {\it the same as} a coherent excitation of a particular oscillator mode, due to a null vector in the current algebra $\exp[i2q\sqrt\nfive\, \cY_\su] = (J^+_{-q})^{q\nfive}$ relating zero modes and oscillators.

It turns out, however, that the remaining constraints do not allow a solution with $w_y=0$ unless the energy $E$ is order one and the $\sltwo$ energy $\msl+\bmsl=O(\Ry)$; these states decouple in the $\Ry\to\infty$ AdS decoupling limit.  In other words, the analogues of giant graviton states in $\ads3\times\bS^3$ cannot expand purely in the $\bS^3$; if they expand in the $\bS^3$, they must also expand in the AdS$_3$.

\subsubsection{Large gauge transformations}
\label{sec:GaugeSpecFlow}

A subtlety arises when considering spectral flow in the null-gauged WZW model~--   
spectral flow in the null direction corresponding to the gauge current is gauge-trivial.  Thus, a linear combination of spectral flows in $\sltwo$ and $\sutwo$ together with shifts in the zero modes $E,n_y, w_y$  amounts to a gauge transformation, and is thus redundant; however this is not true separately for the left and right gaugings, because the $t$ and $y$ directions do not have chiral zero modes.  

Let us bosonize the $\sltwo$ and $\sutwo$ contributions to the null current,
\be
J_\sl^3 = i\sqrt{n_5}\,\partial\cY_\sl 
~,~~~~
J_\su^3 = i\sqrt{n_5}\,\partial\cY_\su ~;
\ee
then the chiral spectral flow operator along the gauge direction is 
\be
\exp[iq\cY] \,\equiv\, \exp\Bigl[iq\sqrt{\nfive}\Bigl( - \cY_\sl+\frac{l_2}{l_1}\,\cY_\su\Bigr)+iq\Bigl(-\frac{l_3}{l_1} \,t + \frac{l_4}{l_1} \,y \Bigr)\Bigr]
\ee
and similarly for the right movers.  
The problem is that these two operators want to shift the zero mode of $t$ differently, but since $t$ is non-compact this is not possible.  This problem does not arise if the gauge group is 
$\bR_{\rm vector}\times U(1)_{\rm axial}$ 
rather than $U(1)_L\times U(1)_R$ -- then the only large gauge transformation corresponds to axial spectral flow, which in our conventions shifts the zero mode of $t$ on the left and right in the same way, as a consequence of the relation $l_3/l_1=r_3/r_1$ (the first of the relations~\eqref{ratios}).

The allowed spectral flow determines the global structure of the sigma
model target space as follows.  If the target space of the sigma model
uses the group manifold $\sltwo$, then the spectral flow parameters
$\wsl,\bwsl$ are allowed to be different, whereas if the target space
involves the universal cover of $\sltwo$, spectral flow is restricted
to $\wsl=\bwsl$.    With gauge group $\bR_{\rm vector}\times U(1)_{\rm
  axial}$, a problem would arise if the $\sltwo$ component of the
target space was the group manifold and not its universal cover; with
only axial spectral flow being gauged, states with different values of
$\wsl-\bwsl$ would constitute distinct physical states, and one
arrives at a nonsensical spectrum.  With a target space involving the universal cover of $\sltwo$, one is restricted to $\wsl=\bwsl$, and the absence of gauged vector spectral flow in the null direction does not lead to pathologies.

Spectral flow by $q$ units in the axial gauge direction is implemented
by the operator $\exp[iq(\cY\tight+\bar\cY)]$, which shifts the zero
modes by:%
\footnote{In discussing the winding sectors, we find it more convenient to use the parameters $s,\bar s$ rather than $\m,\n$, and will do so in what follows.}
\begin{align}
\label{LargeGaugeTransf}
\wsl \to \wsl - q 
~&,~~~
\wsu  \to \wsu  + (2s\tight+1)q
~,~~~
\bwsu \to \bwsu + (2\bar s \tight+ 1)q ~,
\nn\\
E\to E+kR\varrho \,q
~&,~~~~
n_y\to n_y - k\Ry^2\vartheta  \, q
~,~~~~
w_y\to w_y + kq  ~,
\end{align}
where again we have used~\eqref{ratios}.
This set of shifts of zero modes amounts to a large gauge transformation; thus states related by such shifts are gauge-equivalent.  As a consequence, we can choose to work with the representative of each gauge equivalence class having $\wsl=\bwsl=0$ whenever possible, and we shall do so in what follows.

\subsubsection{Brane/flux transitions}
\label{sec:Brane2Flux}

Consider now the sectors of nonzero winding $w_y$.
Gauged axial spectral flow affects the analysis as follows.  There are no non-contractible curves on the universal cover of $\sltwo$, in particular there is no conserved winding around the $\sigma$ circle, and indeed the spectral flow quanta $\wsl,\bwsl$ are not conserved in correlation functions~\cite{Maldacena:2001km}.  On the other hand, the $y$ circle of the WZW model {\it is} non-contractible, with $w_y$ a conserved winding number.  However there are no non-contractible circles in the supertube geometry.  The fact that winding on the $y$ circle is gauge-equivalent to winding on the $\sigma$ circle, and that the latter is non-conserved, allows the gauged WZW model to reproduce the trivial first homology group of the JMaRT spacetime.  

The gauging that relates the $y$ and $\sigma$ winding operates differently in the asymptotic region and the cap, because the $\partial\sigma$ contribution to the $\sltwo$ currents~\eqref{slsucurrents} is exponential in $\rho$, while the $\partial y$ contribution is proportional to $\Ry$.  Thus in the linear dilaton region one is mostly gauging away the $\sltwo$ angular direction, while in the cap one is mostly gauging away the $y$ circle.%
\footnote{Note that this heuristic explains why the local geometry of the cap is $\ads3\times\bS^3$, while asymptotically it has a linear dilaton due to the proper size of the circle which is being gauged away.}
Thus only when the string gets into the cap will the non-conservation of winding become active, because a non-negligible part of the physical winding current starts to come from the $\sltwo$ contribution; this is just what one expects from the effective spacetime picture. 

Note that the shifts in the quantum numbers~\eqref{LargeGaugeTransf} implemented by a large gauge transformation are precisely in proportion to the F1 flux contributions to the energy, momentum and angular momenta~\eqref{eq:E-P}, \eqref{JLandJR} of the background -- we can simply identify $\delta n_1 = q k$.%
\footnote{The energy~\eqref{eq:E-P} agrees with the shift
  $qk\Ry\varrho$ of~\eqref{LargeGaugeTransf} in the supergravity
  approximation and decoupling limit; away from this limit, there are
  $1/\Ry$ and $1/\nfive$ corrections given by~\eqref{eq:NS5decoupledMadm} and \eqref{varrhovartheta}.
}
We can then interpret a process that does not conserve $\sltwo$ winding $\wsl$ by writing the vertex operators of nonzero $\wsl$ as large gauge transformations of vertex operators with $\wsl=0$.  Then the $\wsl=0$ vertex operators involved in the correlator do not by themselves conserve energy, or $y$ circle winding/momentum, or $\bS^3$ angular momentum; the difference is made up by the inserted gauge spectral flow operator(s) implementing a large gauge transformation.  The gauge spectral flow operator acts to screen the charge mismatches (apart from $\sltwo$ winding violation) and allows the correlator to be nonzero.

We thus propose a physical interpretation of the gauge spectral flow operators $\exp[iq(\cY\tight+\bar\cY)]$ as implementing brane/flux transitions wherein the F1 flux $n_1$ in the background changes by the emission or absorption of a corresponding amount of perturbative string winding charge $\delta n_1 = -\delta w_y = -qk$.

This phenomenon was recently investigated in the closely related context of string excitations around global $\ads3\times\bS^3$~\cite{Porrati:2015eha}.  Worldsheet string theory on $\sltwo\times\sutwo\times \cM$ provides a perturbative approximation to near-vacuum states and processes in a spacetime CFT with central charge $c_{\rm ST} = 6n_5n_1$, where $n_1$ is the string winding charge on the $y$ circle.  Again the $\sltwo$ spectral flow sectors describe strings that wind the angular direction of AdS$_3$.  However if we admit vertex operators that create and destroy strings carrying this charge, then either we are not working in a theory with a fixed spacetime central charge, or else the effect of these vertex operators has to be compensated by a change in the number of electric $H_3$ flux quanta in the background, since this flux changes by one unit every time one crosses the worldsheet of the winding string.  This issue was encountered early on in the worldsheet approach to AdS$_3$ string theory~\cite{Giveon:2001up} -- it was noticed that the central charge in the spacetime Virasoro algebra computed via worldsheet vertex operator methods seemed to depend on the correlation function being evaluated, and thus was not actually a c-number.  The resolution of this puzzle came when it was pointed out that calculations involving winding sectors in the $\sltwo$ WZW model take place at fixed {\it chemical potential} for string winding charge, rather than at fixed charge~\cite{Porrati:2015eha}; and that a Legendre transformation is required to ensure that the charge is kept fixed.  When we do this, we are keeping fixed the sum of the flux $n_1$ in the background and the total winding charge $w_y^{\rm tot}$ of perturbative strings.  Correlation functions that do not conserve $\sltwo$ winding $\wsl$ thus describe (after Legendre transformation) processes wherein strings scatter off the background, and in the process some of the background flux is converted to perturbative string winding or vice versa. For a related discussion see~\cite{Giusto:2010gv}. 

Coupled to this issue is one of energy conservation.  When F1 charge is carried by background flux, the corresponding rest energy is ordinarily subtracted out, and spacetime CFT energy is measured relative to the rest energy of the background.%
\footnote{In the Neveu-Schwarz sector of the CFT there is an additional offset $-c/24$ contributed by the vacuum Casimir energy.}
When considering perturbative strings winding the $y$ circle in the gauged WZW model, however, the string energy is explicitly the winding energy $w_y R$ up to small corrections (as one sees from the Virasoro constraints~\eqref{VirConstraints}), naively suggesting that such strings are not allowed in the AdS decoupling limit $R\to\infty$.  The resolution is to adopt a convention that includes the ADM mass of the background, and to impose conservation of the total ADM F1 charge, including that of perturbative strings as well as the background.  This convention is indeed consistent with the zero mode shifts~\eqref{LargeGaugeTransf} generated by large gauge transformations~-- physical states related by large gauge transformations have $E$ differing by the winding energy, which matches precisely the energy (and other charges) associated to the winding quanta being exchanged with the background.

The equivalence relation given by the above shifts~\eqref{LargeGaugeTransf} implies that the winding charge $w_y^{\rm tot}$ of perturbative strings is only conserved modulo $k$.  This seems to be an analogue of the conservation of twisted sector $\bZ_k$ charge in orbifold theories, though once again, we stress that the gauged WZW model is {\it not} a global orbifold.  Nevertheless, it seems that the residue of $w_y$ mod $k$ is unaffected by brane/flux transitions and is a characteristic similar to the magnetic $\bZ_k$ charge that characterizes the twisted sectors of an abelian orbifold.  A qualitative explanation of this feature comes from the description of the supertube ground state in the symmetric product CFT reviewed in Section~\ref{subsec:holographicdescription}.  There, all strands of the background state have length $k$; 
this state admits a $\bZ_k$ symmetry that cyclically permutes the component CFTs making up each strand, which seems to be the analogue in the perturbative spacetime CFT of the $\bZ_k$ symmetry seen here.  Note that this is a symmetry of the particular state being considered, rather than a symmetry of the CFT.  If we now add on top of this a perturbation carried by a strand whose length $\ell$ is not a multiple of $k$, the residue $\ell$ mod $k$ will be conserved by the dynamics.

\subsection{Winding sectors}
\label{sec:winding}

Finally, we examine the spectrum in the sectors of nonzero $w_y$. 
We will make an assumption about the order of terms in the large $\Ry$ expansion, namely that none of the quantum numbers are themselves of order $\Ry$.  In the discussion of supergravity there was a similar issue with scattering states -- in that case the vectorial Virasoro constraint~\eqref{sugraMassShell} placed an order one lower bound~\eqref{Elowerbound} on the energy $E$, and this forced the $\sltwo$ energy $\msl+\bmsl$ to be of order $\Ry$.  This makes sense because these states must have enough energy to reach the AdS boundary, which is at energies of order $\Ry$ times the natural AdS energy gap, which is of order $1/\Ry$.

In the null constraints one has chosen a definite relative sign \eg\ for $l_3$ and $l_4$.  As a consequence, if no other terms compete with the terms involving $E$ and $w_y \Ry$, the null constraints set $E=w_y \Ry$ to have a fixed relative sign which is the same as the sign of the F1 flux in the background.  This is quite different from winding strings in flat spacetime where one determines $E$ by solving the vectorial Virasoro constraint, which gives $E^2=w_y^2 \Ry^2$ as the biggest terms; one then has $E = \pm |w_y| \Ry$, \ie\ the signs of $E$ and $w_y$ are not correlated.  Since the standard field theory conventions are that positive (negative) energy modes are associated to annihilation (creation) operators, we have in this case vertex operators that both create and annihilate strings with either sign of $w_y$.  But in the gauged WZW model, it seems that creation operators introduce perturbative strings with positive energy and positive winding, and annihilation operators remove them; but there are no vertex operators that create/annihilate strings with positive energy and negative winding.

This is as it should be, because as discussed above the perturbative strings are exchanging winding charge with the flux background, and one needs to do a Legendre transform to fix the charge.  When we do this, correlation functions that create a string with winding do so by removing a unit of flux from the background, and similarly add background flux when annihilating a winding string.  All this is consistent with leading order energy conservation if we remember that the string winding contribution to the energy is $(n_1+|w_y|)\Ry$ to leading order if we don't throw away the ADM mass of the background.

Contrast this situation with what one would have with the creation of an antiwound string.  Then charge conservation gives a $\delta n_1$ of the opposite sign, and all of a sudden the string winding energy changed by $2 |w_y| \Ry$.  So there must be an additional term in the energy balance, and indeed there is.  Going back to the null constraint, instead of the $E$ and $w_y \Ry$ terms cancelling to leading order, they add.  We can still solve the constraint, but only if \eg\ one cranks up $\msl+\bmsl$ to enormous values of order $\Ry^2$. 
So there are in principle string/flux annihilation processes described by the gauged WZW model, where an antiwound string annihilates against the flux background, but only at energies of order $\Ry$ above the ground state which is well outside the $O(1/\Ry)$ range of the decoupling limit.


\subsubsection{Bound states}
\label{sec:boundandwound}

Introducing a single fundamental string carrying $w_y>0$ units of background F1 charge, the constraints~\eqref{jmartconstrclean}, \eqref{VirConstraints}
set a threshold energy of order $n_5/\Ry$ above the rest energy for perturbative string states that wind the $y$ circle and are not bound to the cap.  This energy threshold is much lower than the order one energy of scattering states with $w_y=0$, and so the lightest scattering states come from these sectors.  For simplicity, we will look for such states near threshold in the large $\Ry$ limit, and furthermore we assume a large $\nfive$ charge. 

For $\cD^\varepsilon_j$ bound states we have $\msl =
\varepsilon(\jsl+n)$, $\bmsl=\varepsilon(\jsl+\bar n)$ with
$\varepsilon=\pm$. From~\eqref{jmartconstrclean} the axial null
constraint then sets 
\begin{multline}
\label{AxialNullConstraint}
\varepsilon(n-\bar n) \;=\; 
-(2s\tight+1)\Bigl(\msu+\frac\nfive2\wsu\Bigr)
+(2\bar s \tight+1)\Bigl(\bmsu+\frac\nfive2\bwsu\Bigr)\\[5pt]
-k n_y +\frac{s(s\tight+1)-\bar s (\bar s\tight+1)}{k}\, \nfive w_y ~.
\end{multline}
Compared to~\eqref{SugraAxialNullConstraint}, the additional terms are
all integer multiples of $\nfive$.  If these additional terms cancel
among themselves, then the constraint has the same solutions as it had
for the supergravity states;
if not, one or both of $n,\bar n$ are shifted by amounts of order $\nfive$.  We thus set
\begin{equation}
\label{nexpansion}
n = \nfive \ntil + n_s
~~,~~~~
\bar n = \nfive \bntil + \bar n_s
~~,~~~~
n_y = \nfive \nytil + n_{y,s}
\end{equation}
where the ``supergravity quanta'' $n_s,\bar n_s, n_{y,s}$ solve the mod $\nfive$ residue of the constraint, Eq.\;\eqref{SugraAxialNullConstraint}, and 
\begin{equation}
\varepsilon(\ntil-\bntil) = -k \nytil 
-\half(2s\tight+1)\wsu
+\half(2\bar s \tight+1)\bwsu
+\frac{s(s\tight+1)-\bar s (\bar s\tight+1)}{k}\,w_y ~.
\end{equation}

We work in the gauge where the
$SL(2,\mathbb{R})$ spectral flows are set to zero: $\wsl = \bwsl = 0$. 
For states in the discrete series $\cD^\pm_j$ of $\sltwo$, the sum of
the Virasoro constraints~\eqref{VirConstraints} determines one relation between the energy $E$ and the $\sltwo$ spin
$\jsl$; then the sum of the null constraints~\eqref{jmartconstrclean}
determines another such relation.
In addition, the axial null constraint~\eqref{AxialNullConstraint} and
the $L_0 - \bar L_0$ constraint 
\begin{equation}
0 = n_y w_y + \bigl(\msu\wsu - \bmsu\bwsu\big)+ \frac{n_5}{4}(\wsu^2-\bwsu^2) + \NL-\NR 
\end{equation}
are Diophantine relations among the quanta.  The solution to the axial null constraint was discussed above; the axial Virasoro constraint typically requires string oscillator modes to be excited by amounts of order $\nfive$ if $\wsu\ne\pm\bwsu$.  Note that a change $\delta n_y=1$ must typically be compensated by a change $\delta(\NR\tight-\NL)=w_y$ -- the string winding fractionates oscillator momentum by a factor of the winding number.  As a result, winding strings (when present) carry more entropy than supergravity modes for a given energy.

At large $\Ry$ and at leading order in $\nfive$, for both discrete series representations $\cD^\varepsilon_j$ one can approximate
\begin{align}
\label{BoundStateQnums}
\varepsilon\,\ntot &\sim 
{\nfive}\biggl[
\frac{\bigl(s(s\tight+1)+\bar s(\bar s\tight+1)\bigr) w_y}{k}
- \frac{(2s\tight+1)}2 (\Msu\tight+\wsu) - \frac{(2\bar s\tight+1)}2 (\bMsu\tight+\bwsu)
\nn\\
& \hskip 1cm 
- \varepsilon\Jsl +\frac{k}{4w_y}\Bigl( \wsu(\wsu\tight+2\Msu) +
  \bwsu(\bwsu\tight+2\bMsu) + 4\Ntot - 2\Jsl^2 + 2\Jsu^2\Bigr) 
\biggr] \, ,
\nn\\[.5cm]
\delta E & \sim \frac{\nfive}{4 \Ry w_y}
\Bigl[ \wsu(\wsu\tight+2\Msu) + \bwsu(\bwsu\tight+2\bMsu) + 4\Ntot -
           2\Jsl^2 + 2\Jsu^2\Bigr] \, ,
\nn\\[.5cm]
\Jsl & \sim \frac{\varepsilon}{k} \Bigl[ -w_y + \sqrt{ \Delta } \;\Bigr]\, ,
\\[.5cm]
\Delta &=  \frac12 \Bigl[
k^2\bigl( 2\Jsu^2 -\Msu^2-\bMsu^2 + 4 \Ntot\bigr) - \varepsilon \,4k\,\ntot w_y 
\nn\\
& \hskip 1cm  + \bigl((2s\tight+1)w_y-k(\Msu+\wsu)\bigr)^2 
+ \bigl((2\bar s\tight+1)w_y-k(\bMsu+\bwsu)\bigr)^2
\Bigr] \, ,
\nn
\end{align}
where $\delta E$ is the energy relative to the rest mass $w_y\Ry$, and we define quantum numbers as fractions of $\nfive$ via
\begin{align}
\msu=\frac\nfive2 \Msu
~,~~
\bmsu=\frac\nfive2 \bMsu
&~,~~
\jsl = \frac\nfive 2 \Jsl
~,~~
\jsu = \frac\nfive2 \Jsu~,
\\
n+\bar n = \nfive\,\ntot
~,~~&
\NL+\NR = \nfive \Ntot
\nn
\end{align}
(thus unitarity imposes $\Jsu,\Jsl\lesssim 1$).
Note that we have chosen a particular branch of the square root in
$\Jsl$; for $\cD^+_j$, the other choice leads to a value outside the
unitarity bound, while for $\cD^-_j$ the resulting state can be mapped back to the other branch via field identifications (see the discussion at the end of Appendix~\ref{app:WZWmodels}).

The relative energy $\delta E$ is minimized by maximizing $\Jsl$, and for fixed values of the angular momenta this occurs for $\ntot=0$ (for both signs of $\varepsilon$).  If we then demand that these perturbative winding strings carry approximately the same momentum and angular momenta as the background,
\be
\label{BkgdQuantNums}
\wsu\tight+\Msu \sim \frac{2s\tight+1}{k} w_y
\,,~~~
\bwsu\tight+\bMsu \sim \frac{2\bar s\tight+1}{k} w_y
\,,~~~
\nytil \tight+ \frac{n_{y,s}}{\nfive} \sim \frac{\bar s(\bar s\tight+1)-s(s\tight+1)}{k^2} w_y \,,
\ee
then the lower bound on the relative energy reduces to (setting $\ntot=0$)
\be
\label{Eleast}
\delta E \gtrsim \frac{\nfive}{k^2\Ry} \biggl[
\bigl(s(s\tight+1) + \bar s(\bar s\tight+1)\bigr) w_y
+ \sqrt{ k^2\Bigl(\Jsu^2 - \hf\Msu^2-\hf\bMsu^2 +2\Ntot \Bigr) 
}  \;\biggr] \,.
\ee
Thus the energy per unit F1 winding is larger than that of the background (compare~\eqref{eq:E-P} with with $\delta n_1=w_y$).

A consequence of this result is that the background cannot decay via the emission of perturbative strings -- there is no decay channel whereby dissolved F1 flux releases free energy by turning into some number of perturbative strings carrying the same quantum numbers.  
On a given string, worldsheet spectral flow is the least energetic way of adding angular momentum; one way to see this is to factor the worldsheet CFT into the U(1) of the angular momentum current being flowed, and a coset CFT that is neutral under the current.  The quadratic energy of spectral flow is that of the exponential of the bosonized current that carries the angular momentum charge; that is then correlated with a coset contribution which is strictly positive, by unitarity of the coset CFT.
One can ask whether there is an energetic advantage to redistributing the angular momentum among multistring states.  For instance, given two strings of winding $w_y$, one might take a fraction $\epsilon$ of angular momentum from one and put it on the other; but since the energy cost is quadratic in the angular momentum, the energy increases by a factor $1\tight+\epsilon^2$.  Alternatively, one can ask whether there is an energetic advantage to putting spectrally flowed angular momentum on strings of larger or smaller winding.  But a spectral flow energy of two strings carrying angular momentum $J=\alpha w_y$ each amounts to $E_{\rm specflow}=2\cdot{\alpha^2w_y}/4 =(2J)^2/(4 (2w_y)) $, which is the same as the spectral flow energy of a single string of winding $2w_y$ carrying angular momentum $2J$; thus there is no advantage to distributing the spectral flow energy among strings of any particular length (longer or shorter).  Similarly, from the constraints~\eqref{jmartconstrclean} one sees that partitioning the angular momenta $(2s+1)w_y/k$, $(2\bar s+1)w_y/k$ among any number of strings with winding zero, the energetic cost attributed to that angular momentum remains the same.
Thus spectral flow is the least energetic way of putting angular momentum on the system.  Supergravity quanta carrying the same angular momenta cost as much or more energy, and redistributing the angular momenta among the winding strings also does not result in a lower energy state that would free up some energy and initiate a decay of the background.

Deeply bound discrete series winding states provide the closest possible analogues of the twisted-sector states of non-supersymmetric orbifolds. Whereas the ground states in such non-supersymmetric orbifold twist sectors can be tachyonic, it seems that there are no similar perturbative instabilities in the supertube backgrounds of Section~\ref{sec:fivebranedecouplinglimit}.
There is no paradox here, because the gauged WZW model is not a global orbifold.

If however the angular momentum per unit winding is less than that of the background, the energy~\eqref{Eleast} of the string can be less than that of the background, for given amount of winding, while solving all the constraints.  The winding string is then deeply bound to the background, and one can excite it with a large number of oscillator excitations before exceeding the bound state threshold.  Furthermore, the winding string energy~\eqref{BoundStateQnums} reveals that the energy cost of oscillator excitations is fractionated by the winding charge~$w_y$.

Note that both positive and negative discrete series representations $\cD^\varepsilon_j$ can occur, even for these strings carrying charges proportionate to the background.  Restoring $\ntot$, the $\sltwo$ spin $\Jsl$ for these background-proportionate strings is
\be
\label{Jsl_bkgd_chg}
\Jsl =  \frac{\varepsilon}{k} \biggl[
 - w_y
+ \sqrt{ k^2\Bigl(\Jsu^2 - \hf\Msu^2-\hf\bMsu^2 +2\Ntot \Bigr) 
-2\varepsilon k\ntot w_y
}  \;\biggr] \,.
\ee
For given quantum numbers, typically only one of these is in the unitary range $0<\Jsl<1$.  When $\Msu=\Jsu$, the first term under the square root is $2\bigl(s(s\tight+1)-\bar s(\bar s\tight+1)\bigr) w_y^2$; thus if we consider the ground state $\Ntot=\ntot=0$, it must lie in $\cD^+_j$ for $s>\bar s>0$.  On the other hand, if we set $s=\bar s>0$, then the ground state lies in $\cD^-_j$ (and as for super-radiant supergravity modes, we expect these winding states to become super-radiant instabilities when the fivebrane throat is glued back onto asymptotically flat spacetime).%
\footnote{It seems that $s=\bar s$, and backgrounds related to it by integer spectral flow in spacetime, are the only examples exhibiting this super-radiant phenomenon in winding sectors.}
As one increases \eg\ the oscillator excitation $\Ntot$ above the ground state, $\Jsl$ will eventually be pushed outside the unitary range  unless one increases the cap energy $\ntot$ as well.  This has the curious feature that the cap energy $\msl+\bmsl$ is becoming more and more negative in the process.

The wound strings considered here are analogues of the supertube probes employed in~\cite{Bena:2008dw,Bena:2008nh,Bena:2010gg} in the search for highly entropic horizonless configurations.  The idea in these works was that the ``effective'' charges of the supertube are determined by its local environment in the cap rather than the charges at infinity, and so the scope for exciting oscillators on the supertube might be enhanced relative to a supertube in isolation.  It was found that this enhancement favored a string counter-rotating relative to the background.\footnote{We also note in passing the studies of probe supertube ergoregions in~\cite{Chowdhury:2013ema}.}
Along these lines, consider a wound string in the $\cD^+_j$ representation carrying momenta near the background values
\be
\wsu\tight+\Msu \sim \frac{2s\tight+1}{k} w_y + \delta\wsu
~,~~~~
\bwsu\tight+\bMsu \sim \frac{2\bar s\tight+1}{k} w_y + \delta\bwsu
~;
\ee
the string energy and $\sltwo$ spin are 
\begin{align}
\Jsl  &\sim  \frac{1}{k}\Bigl[ -w_y + \sqrt{\Delta }\; \Bigr] \,,
\nn\\[.3cm]
\delta E & \sim \frac{\nfive}{k^2\Ry} \Bigl[ \frac{k}2\bigl( (2s\tight+1)\delta\wsu +(2\bar s\tight+1)\delta\bwsu + 2\ntot\bigr)
+ \bigl(s(s\tight+1)+\bar s(\bar s\tight+1)\bigr) w_y +\sqrt\Delta \;\Bigr] \,,
\nn\\[.3cm]
\Delta&= \half k^2\bigl( \delta\wsu^2+\delta\bwsu^2 +2\Jsu^2-\Msu^2-\bMsu^2 + 4\Ntot\bigr)-2k\,\ntot w_y \,.
\end{align}
Naively, for $s,\bar s>1$ one would want to make $\delta\wsu,\delta\bwsu$ as negative as possible, however one is constrained by having the $\sltwo$ spin $\Jsl$ in the unitary range.  As their magnitude is increased, one must increase $\ntot$ in order to keep $\Delta$ sufficiently small.
Therefore let us set 
\be
\ntot=\frac{k(\delta\wsu^2+\delta\bwsu^2+4\Ntot)}{4w_y} + \delta\ntot ~;
\ee
the energy becomes
\begin{align}
\delta E  &\sim \frac{\nfive}{4w_y\Ry} \biggl[  \delta\wsu \Bigl( \delta\wsu+2(2s\tight+1)\frac{w_y}k\Bigr) 
+ \delta\bwsu\Bigl(\delta\bwsu + 2(2\bar s\tight+1)\frac{w_y}{k}\Bigr) + 4\Ntot
\nn\\
& \hskip 2cm
+ 4\,\delta\ntot \frac{w_y}{k}   
+ 4\Bigl(s(s\tight+1)+\bar s(\bar s\tight+1)\Bigr) \frac{w_y^2}{k^2} +4\frac{w_y}{k}\sqrt{\delta\Delta} \;\biggr]\,,
\\[.3cm]
\delta\Delta & = \Jsu^2-\half \Msu^2-\half \bMsu^2 - 2 \,\delta\ntot \frac{w_y}{k} ~.
\nn
\end{align}
The second line in the expression for the energy is approximately proportional to the background energy for the given charges; it is the first line where one has some interesting scope for adjustment.  By allowing $\delta\wsu,\delta\bwsu$ to be negative up to amounts of order $(2s\tight+1)\frac{w_y}k,(2\bar s\tight+1)\frac{w_y}k$, respectively, one creates a ``reservoir'' of negative energy relative to the background%
\footnote{Which is largest if the string carries no angular momentum at all.} 
that can be filled up with oscillator energy $\Ntot$.  Thus the string lags the background rotation, and that frees up some energy for oscillator excitation.  Of course, the system as a whole carries less angular momentum per unit winding than a state where all winding strings carry angular momentum proportional to the background; but if the system is able to shed some angular momentum while retaining some of the corresponding energy, \eg\ through super-radiant emission, such states may be the entropically preferred direction in which the system evolves, at least initially.

\subsubsection{Scattering states}

We now consider states that belong to continuous series
representations of $\sltwo$, whose wavefunctions are plane
waves in the radial direction and thus represent scattering
states.  For these representations, we analytically continue $\nu \to i \nu$ with respect to \eq{eq:jslsugra}  and write the $\sltwo$ spin as (c.f.~\eq{eq:cont-series-app})
\be
\jsl \,=\, \frac12 + i \frac{\nu}{2} \,.
\ee
We assume that $\msu= \bmsu=0$ for simplicity. 

A simple solution to all the constraints sets the various winding
numbers to be proportional to the background values (taking the winding $w_y$ to be a multiple of $k$):%
\footnote{This restriction is not necessary -- there are scattering states in all winding sectors, which have the same structure, but the expressions are more complicated.  Once again, at large $\nfive$ one can perform an analysis along the lines of the previous subsection, taking $\msu/\nfive,\bmsu/\nfive$ as fractional parts of $\wsu,\bwsu$.}
\begin{align}
\label{bkgdvals}
\jsu=0
~,&~~~
\wsu=\frac{2s+1}{k} w_y
~,~~~
\bwsu=  \frac{2\bar s+1}{k} w_y
~,~~~
\nu^2 = \frac{\nfive^2 w_y^2}{k^2}-1 \;, \\[.1cm]
& \msl=\bmsl=0 
~,~~~
n_y =\frac{ \bar s(\bar s\tight+1)-s(s\tight+1)}{k^2}n_5 w_y ~,
\nn
\end{align}
with the energy slightly above threshold
\begin{equation}
E - w_y \Ry = \frac{\nfive\bigl[ s(s+1)+\bar s(\bar s+1) +1\bigr] w_y}{k^2\Ry} = \frac{\nfive\bigl( \m^2+\n^2 +1\bigr) w_y}{2k^2\Ry}  
\end{equation}
(again compare~\eqref{eq:E-P}, with $\delta n_1=w_y$).  

More generally, solving the vectorial null constraint and plugging
into the vectorial Virasoro constraint, in the large $\Ry$ expansion
one finds at leading order
\begin{align}
\label{lambdaSoln}
k^2\nu^2 \;=\; 
-k^2(2\jsu\!+\!1)^2 &- 2\nfive k^2(\NL\!+\!\NR) + 2\nfive k(\msl+\bmsl)w_y\ 
\\[.3cm]
-\frac{n_5^{\,2}}{2} &\Bigl[ \bigl((2s \tight +1)w_y-k\wsu\bigr)^2 +
                       \bigl((2\bar s \tight +1)w_y-k\bwsu\bigr)^2 -
                       2w_y^2 \Bigr] \, ,  \nn
\end{align}
Note that typically all the terms on the RHS are negative, except for
the $\msl+\bmsl$ terms.  It seems we have two options: the first is to
have $w_y$ a multiple of $k$, and adjust $\wsu,\bwsu$ to set the first
two squares on the second line to vanish, which leads back
to~\eqref{bkgdvals}; the second is to adjust the $(\msl+\bmsl)$ term
of the first line to be sufficiently positive to compensate all the
other terms.  The first option solves the constraints without turning on a large amount of $\sltwo$ energy, while the second relies on such an energy of order $\nfive$; these solutions with large positive $\sltwo$ energy allow states where modes on the winding string carry fractionated momenta, as we now show.

The energy of scattering states is determined by solving the vectorial
null constraint:
\begin{align}
E-w_y \Ry = \frac{1}{2k^2\Ry}\Bigl[2 k(\msl+\bmsl) 
+ \nfive k&\bigl((2s+1)\wsu + (2\bar s +1)\bwsu\bigr) \nn \\ &-
2n_5\bigl(s(s+1)+\bar s (\bar s+1)\bigr)w_y\Bigr] \,.
\end{align}
Generically $(\msl+\bmsl)$ must be positive and of order $\nfive$ in order to have a solution to~\eqref{lambdaSoln}; the threshold value is obtained approximately by setting $\nu=0$.
Substituting into the above yields an approximate lower bound on the
energy of scattering states
\be \label{eq:bound-scat}
\Bigl(E - w_y\Ry\Bigr) \,\gtrsim\, \frac{1}{w_y\Ry} \biggl[ \frac{(2\jsu\tight+1)^2 }{2\nfive}
+ (\NL+\NR) + \frac{n_5}4 (\wsu^2+\bwsu^2) \biggr]
\,.
\ee
The quantity in square brackets is the energy above the rest mass of the macroscopic winding string, again fractionated by the winding $w_y$.

While perturbative winding strings fractionate momentum by the amount of F1 winding charge they carry, they fall well short of the fractionation afforded by the spacetime CFT in the black hole regime, which comes with an additional factor of $\nfive$ (see for instance the analysis~\cite{Giveon:2015raa} of the elliptic genus in little string theory).  This factor of $\nfive$ reflects the fact that little strings rather than fundamental strings are the carriers of black hole entropy.

\section{Two-point correlation function}
\label{sec:Correlators}

In a separable wave equation, bound states are associated to poles in the reflection coefficient of the radial scattering problem.  In the present context, this reflection coefficient is a component of the two-point function of string vertex operators.  For scattering states, this same reflection coefficient is a phase shift which encodes the time delay encountered by strings scattering off the cap in the geometry.

We can construct a natural set of observables in the gauged WZW model
from vertex operators in the group $\mathcal{G}$, with quantum numbers
constrained by \eqref{jmartconstrclean}.  The vertex operators decompose into contributions from each factor of the $\cG$ WZW model,
\begin{equation}
V_{\mathcal{G}} = \Phi_{\jsl,\msl,\bmsl}^{\sl}
\Phi_{\jsu,\msu,\bmsu}^{\su} e^{i E t} e^{i P y} \,,
\end{equation}
where we suppress the polarization structure, ghost contributions, and so on, that play no essential role in what follows.  We ignore the possibility of momentum/winding on $\cM$, though this is not essential.

The modes $\Phi_{\jsl,\msl,\bmsl}^{\sl}$, $\Phi_{\jsu,\msu,\bmsu}^{\su}$ are superconformal primary fields, and thus are string versions of eigenfunctions of the Laplacian on
AdS$_3$ and $\bS^3$ respectively.
The non-trivial part of string correlators comes from the AdS$_3$
factor.  Their two-point functions are (see for instance~\cite{Giveon:1999px})
\begin{align}\label{eq:PhiPhicorrelator}
\langle \Phi_{\jsl; \msl, \bmsl}^{\sl} \Phi_{\jsl; -\msl , -\bmsl}^{\sl} \rangle &=
n_5\, \mub^{2\jsl-1}\,
\frac{\Gamma (1-\frac{2\jsl-1}{n_5})}{\Gamma (\frac{2\jsl-1}{n_5})}  \\[.3cm]
&\hskip 1cm
\times \frac{\Gamma (-2 \jsl +1)\Gamma (\jsl-\msl)\Gamma (\jsl+\bmsl)}{\Gamma (2\jsl-1) \Gamma (-\jsl -\msl+1)\Gamma (-\jsl+\bmsl+1)} \,. \nn
\end{align} 
Ordinarily in CFT one chooses the two-point function to have unit normalization by a choice of field redefinition, however here the two-point amplitude represents a reflection amplitude for scattering in the non-compact $\sltwo$ geometry, and thus has non-trivial momentum-dependent information.  Similarly, for the $\sutwo$ CFT we choose the normalization
\be
\langle \Phi_{\jsu; \msu, \bmsu}^{\su} \Phi_{\jsu; -\msu , -\bmsu}^{\su} \rangle =
\frac{\Gamma (\frac{2\jsu+1}{n_5})}{\Gamma (1-\frac{2\jsu+1}{n_5})}   \,.
\ee
This choice simultaneously removes certain ``leg-pole'' factors from the $\sutwo$ three-point function, and makes the overall string two-point function agree with that of the spacetime CFT in cases that have been checked~\cite{Galliani:2016cai}, without the use of momentum-dependent rescalings of the vertex operators of the sort used in~\cite{Gaberdiel:2007vu}.

This two-point function has a series of poles at
\begin{equation}
\msl = \jsl + n \,, \qquad n \in \mathbb{N}
\end{equation}
or 
\begin{equation}
\label{eq:sutwo 2pt}
\bmsl = -\jsl - \bar n \,, \qquad \bar n \in \mathbb{N} 
\end{equation}
which correspond to the spectrum of bound states in the cap.
This structure was used in~\cite{Aharony:2004xn} to study correlation functions in little string theory (LST).  Since the supertube background is a particular background of LST on $\bS^1\times \bT^4$ (or $K3$) in the superselection sector with $n_1$ units of F1 charge, the same considerations apply here.

For supergravity modes in the AdS$_3$ limit, one has $\jsl=\jsu+1$, see Eq.\;\eqref{D+sugraconstr}.  Then the gamma function factors in the first line of~\eqref{eq:PhiPhicorrelator} cancel against those of~\eqref{eq:sutwo 2pt},
and the two-point function reduces to
\begin{equation}\label{eq:PhiPhilargen}
\langle \Phi_{\jsl; \msl, \bmsl}^{\sl} \Phi_{\jsl; -\msl , -\bmsl}^{\sl} \rangle  \sim
\mub^{2\jsl-1}\,  \frac{\Gamma (-2 \jsl +1)\Gamma (\jsl-\msl)
  \Gamma (\jsl+\bmsl)}{\Gamma (2\jsl-1) \Gamma (-\jsl -\msl+1)\Gamma (-\jsl+\bmsl+1)}   \ .
\end{equation}
More generally, in the limit $n_5 \rightarrow \infty$ at fixed $j$, the $\sutwo$ two-point function~\eqref{eq:sutwo 2pt} and the first line of the $\sltwo$ two-point function~\eqref{eq:PhiPhicorrelator} tend to a constant, again yielding the above result.
In this limit, we should expect to recover the result of the
supergravity approximation; indeed this expression agrees with  
the ratio of wavefunction asymptotics~\eqref{wavefnasymp} with the substitutions~\eqref{nullconstrs}, \eqref{eq:jslsugra}.  For scattering states, the ratios of gamma functions in the $\sltwo$ correlator contribute phases, and the factor in the first line no longer cancels against the $\sutwo$ factor~\eqref{eq:sutwo 2pt}.  The phase shift encodes the time delay for probes to scatter off the cap and return to infinity; with a hard cap at some fixed redshift in the smeared classical geometry of the supergravity approximation, all probes at high energy will have the same time delay (backreaction decouples for a small-amplitude probe in the classical limit).

The ratio of gamma functions on the first line of~\eqref{eq:PhiPhicorrelator} encodes very interesting stringy physics of scattering states in the $\sltwo/\uone$ cigar coset~\cite{Giveon:2015cma,Giveon:2016dxe}.  Writing $\jsl = \frac12(1+i\nu)$, the stringy gamma functions in the first line of~\eqref{eq:PhiPhicorrelator} give an additional phase shift~\cite{Giveon:2015cma}
\be
\label{phaseshift}
\delta \simeq \frac 2{\nfive} \nu\log\nu \,.
\ee
This phase shift was interpreted using the FZZ duality between the $\sltwo/U(1)$ gauged WZW model and $\cN=2$ Liouville theory~\cite{Giveon:1999px,Hori:2001ax} as arising from a saddle point in the $\cN=2$ Liouville path integral, where the time delay is realized as the increasing time it takes a probe to climb the exponential Liouville wall and fall back down, as the radial momentum is cranked up.

In~\cite{Giveon:2015cma}, the phase shift~\eqref{phaseshift} was considered in the context of the Euclidean geometry of black NS5-branes, where the $\sltwo/\uone$ ``cigar'' coset describes string dynamics in the radial and Euclidean time directions.  The additional time delay was interpreted in terms of strings propagating {\it beyond} the tip of the cigar~-- that somehow perturbative strings could see ``beyond the horizon'' of black NS5-branes.  In the present context, the same coset model describes part of the transverse space of the fivebranes, and the $\cN=2$ Liouville potential was interpreted in~\cite{Martinec:2017ztd} as the stringy replacement for a near-source structure of separated fivebranes.  More precisely, in the NS5-F1 frame one has the T-dual of separated fivebranes.  
The T-dual geometry of separated fivebranes, and the associated breaking of $U(1)$ rotational symmetry of the smeared geometry to the $\bZ_{n_5}$ symmetry of localized sources, was related in~\cite{Gregory:1997te} to instanton effects in the worldsheet sigma model.  However, that analysis takes place in a regime where the fivebrane separation is much larger than the string scale, whereas in the supertube background, the source separation is much {\it smaller} than the string scale (though much larger than the Planck scale).  A geometrical T-dual picture of the near-source structure is supplanted by the tachyon condensate of the $\cN=2$ Liouville theory (note that $\nfive$ only appears nontrivially in the Liouville-associated contributions to the phase shift).  But that is to be expected~-- when curvatures approach the string scale, there can be non-trivial mixing between string modes; indeed, one may regard the FZZ duality as a non-compact version of the Calabi-Yau/Landau-Ginsburg correspondence~\cite{Martinec:1988zu,Greene:1988ut,Witten:1993yc} in which geometry is supplanted by a tachyon condensate in regimes where the curvature is stringy.

From a fivebrane perspective, the cap in the supertube geometry comes because
the fivebranes are forced to separate slightly onto their Coulomb
branch by the angular momentum they carry.  While $\nfive\ge2$
coincident fivebranes generate a geometrical linear dilaton throat
that strings can propagate down, there is no perturbative worldsheet
description of individual isolated fivebranes; thus when a
perturbative string travels radially down the throat far enough that
it should start resolving the individual fivebrane sources, and
getting close enough to the sources that the string coupling should
become large, a low-energy probe of this sort instead meets the cap in the effective geometry and is reflected back.  In the smeared geometry, the running of the dilaton cuts off at the attractor value $Q_1=Q_5$ set by the background charges, \ie\
\be
\gstr^2 = \frac{\nfive}{\none } V_4 \;,
\ee 
and one may regard this effective string coupling as a proxy for how close to the fivebrane sources the probe is able to reach.  At sufficiently large radial momentum, however, strings begin to explore even closer to the strong coupling region near fivebrane sources -- the further into the Liouville wall a probe travels, the larger the effective string coupling it sees (since the Liouville description still has a linear dilaton even in the cap region).  Thus the structure ``beyond the cap'' proposed in~\cite{Giveon:2015cma} may be thought of in the present context as a feature of the stringy substructure of the supertube source, closer to the fivebrane source than low-energy probes can reach.  Probes which can see this substructure provide a diagnostic of the strong-coupling physics that lurks there.

Two classes of scattering states which can access this substructure were discussed above.  Supergravity modes in the continuous series (discussed in Section~\ref{sec:supergravity}) have energies and radial momenta $|\jsl|>\sqrt\nfive$ of order $\Ry^0$; these states decouple in the AdS decoupling limit, which retains only energies that scale as $\Ry^{-1}$.  In addition, winding states have a useful role to play.  Naively their energies are of order $w_y \Ry$, however this energy cost has already been paid for by the change in background flux, as discussed above, and should not be counted against them.  The residual energy can be small, of order $\Ry^{-1}$, while the radial momentum is large relative to the threshold for stringy effects if $w_y\gg k$; see Eqs.\;\eq{lambdaSoln} and \eq{eq:bound-scat}.  In other words, winding strings which are nearly BPS relative to the background can be light relative to the overall mass of the background, and so remain part of the spectrum in the AdS decoupling limit; however they also have a large inertia, and thus serve to probe the stringy aspects of the supertube such as structure ``beyond the cap''.

\vskip 1cm
\section*{Acknowledgements}

We thank \'{O}scar Dias
and Andrea Galliani
for useful discussions.
The work of EJM and SM is supported in part by DOE grant DE-SC0009924, as well as a FACCTS collaboration grant. The work of DT is supported by a Royal Society Tata University Research Fellowship. For hospitality during the course of this work, we thank the Centro de Ciencias de Benasque, and DT thanks the Enrico Fermi Institute, University of Chicago.


\vskip 1cm
\appendix


\section{Ergoregions}
\label{app:ergoregion}

In this appendix we review the fact that the asymptotically flat JMaRT geometry has an ergoregion for any choice of asymptotically timelike Killing vector field, while in the AdS decoupling limit, there is a globally timelike Killing vector field~\cite{Jejjala:2005yu}. We then show that in the NS5 decoupling limit, there is also a globally timelike Killing vector field. This indicates that the ergoregion instability of the asymptotically flat solutions is not present in the decoupled solutions. 

In spacetimes with isometries corresponding to spatial directions which have finite size asymptotically, there is not a unique asymptotically timelike Killing vector field, and so there is no preferred definition of an ergoregion; rather there are different ergoregions corresponding to different choices of asymptotically timelike Killing vector fields (for related discussions see~\cite{Hawking:1999dp,Pelavas:2005}).

The fivebrane and AdS decoupling limits taken in Sections \ref{sec:fivebranedecouplinglimit} and \ref{sec:AdS} result in solutions with different asymptotics. In the full asymptotically flat solutions, only the $y$ circle is finite size at asymptotic infinity. In the NS5 decoupling limit, the $y$ circle and S$^3$ are both finite size in the resulting asymptotically linear-dilaton solution. In the AdS decoupling limit, only the S$^3$ is finite size in the asymptotically AdS$_3 \times$S$^3$ solution.

In the asymptotically flat JMaRT solutions, the most general Killing vector field that is causal (i.e. timelike or null) at infinity is given by
\bea
\d_t + v^y \d_y
\eea
for constant $v^y$ with $|v^y| \le 1$. For any value of $v^y$ this Killing vector field becomes spacelike in the interior of the solutions~\cite{Jejjala:2005yu}.

In the asymptotically $\ads3 \times$S$^3$ spacetimes, the Killing vector field $\d_t$ at fixed $\tilde\psi$, $\tilde\phi$ defined in \eq{AdS3variables} is globally timelike~\cite{Jejjala:2005yu}, as can be seen from the metric \eq{eq:ads}.

It remains to investigate the asymptotically linear-dilaton spacetimes, where the most general Killing vector field that is causal at infinity is given by
\bea \label{eq:norm-KV-NS5}
\d_t + v^y \d_y + \frac{v^\psi}{\sqrt{Q_5}} \d_\psi + \frac{v^\phi}{\sqrt{Q_5}} \d_\phi
\eea
for constants $v^y$, $v^\psi$, $v^\phi$ such that for all $\theta\in[0,\pi/2]$,
\bea
(v^y)^2 + \cos^2\theta \, (v^\psi)^2 +\sin^2\theta \,(v^\phi)^2 \;\le\; 1 \,.
\eea
One can verify that the following components give rise to a globally timelike Killing vector field in the NS5 decoupling limit:
\bea
v_y \;=\;  -\frac{\sinh 2 \xi}{\cosh 2 \xi + \cosh 2\zeta }   \,, \qquad 
v_\psi \;=\; - v_\phi \;=\;  -\frac{2 \cosh \xi \cosh \zeta}{\cosh 2\xi + \cosh 2 \zeta}   \,. 
\eea

\section{Current algebra properties}
\label{app:WZWmodels}


\subsection{\texorpdfstring{$\sutwo$}{}}
\label{sec:sutwocft}

The supersymmetric $\sutwo$ level $\nfive$ current algebra consists of currents $J_\su^a$ and their fermionic superpartners $\psi_\su^a$ having the OPE structure
\begin{align}
J_\su^a(z)\,J_\su^b(0) &\sim \frac{\frac12 \nfive \, \delta^{ab}}{z^2} + \frac{i\epsilon^{abc} J^\su_c(0)}{z}
\nn\\
J_\su^a(z)\, \psi_\su^b(0) &\sim i\epsilon^{abc}\frac{\psi_c^\su(0)}{z}
\\
\psi_\su^a(z)\, \psi_\su^b(0) &\sim \frac{\delta^{ab}}{z}
\nn
\end{align}
with the Killing metric $\delta^{ab} = {\rm diag}(+1,+1,+1)$.  One can define a set of ``bosonic'' $\sutwo$ level $\nfivetil=\nfive-2$ currents $j_\su^a$ that commute with the fermions,
\be
j_\su^a = J_\su^a + \frac i2\epsilon^{abc}\psi^\su_b \psi^\su_c  ~.
\ee
The primary fields $\Phihat_{\jsu \msu \bmsu}^{\su}$ of the current algebra have conformal dimensions
\be
h = \bar h = \frac{\jsu(\jsu+1)}{\nfive} ~;
\ee
unitarity restricts the allowed spins $\jsu$ of the underlying bosonic current algebra to the allowed range
\be
\label{su2 reps app}
\jsu = 0,\frac12,\dots,\frac{\nfive}{2} -1 ~.
\ee

The primary operators of the supersymmetric theory are built by combining primaries $\Phi^\su_{\jsu \msu \bmsu}$ of the level $\nfivetil$ bosonic current algebra with level two primaries of the fermions (namely the identity operator $\One$, spin operator $\Sigma$, and the fermions themselves).  The highest weight fields are of three types:
\be
\label{su2hwr}
\Phihat^\su_{jjj} = \Phi^\su_{jjj}
~~,~~~~
\Phihat^\su_{j+\half,j+\half,j+\half} = \Phi^\su_{jjj}\Sigma_{++}
~~,~~~~
\Phihat^\su_{j+1,j+1,j+1} = \Phi^\su_{jjj}\psi^+\bar\psi^+
~~,
\ee
with the remaining operators of the zero mode multiplet obtained through the action of the zero modes of the total current $J^-_\su$.  In building massless string states, one uses the purely bosonic highest weight operator for NS sector states whose polarization does not lie along $\sutwo$, and the highest weight operator with a fermion attached when the polarization does lie along $\sutwo$; Ramond sector operators involve $\Sigma$, with the various spinor polarizations reached through the action of the zero modes of $\psi^-_\su,\bar\psi^-_\su$.  In the type II string, the choice of fermion decoration is independent on left and right, and there is a chiral GSO projection onto odd total fermion number in the matter sector.

These operators have a {\it parafermion} decomposition%
\footnote{Our notation here largely follows~\cite{Martinec:2001cf}, see also~\cite{Giveon:2015raa}, except that we work in conventions where $\alpha'=1$, so that T-duality is $R\to 1/R$, instead of the convention $\alpha'=2$ of those works.}
under the current $J_\su^3$%
~\cite{Fateev:1985mm,Gepner:1986hr,Gepner:1987qi}
obtained by extracting the dependence on $J^3_\su, \bar J^3_\su$.  To this end, one bosonizes the currents
\begin{align}
j^3_\su=i\sqrt{\nfivetil}\,\partial Y_\su 
~&,~~~~
\psi^+_\su \psi^-_\su = i\sqrt2\,\partial H_\su ~,
\nn\\
J^3_\su=i\sqrt{\nfive}\,\partial \cY _\su
~&,~~~~
J_\cR^\su = \frac\nfivetil\nfive \psi^+_\su\psi^-_\su - \frac2\nfive j_3^\su
= i\sqrt{\frac{2\nfivetil}{\nfive}} \, \partial \cH_\su ~,
\end{align}
and similarly for the right-movers.  The current $J^3_\su$ forms a $U(1)$ supermultiplet with the fermion $\psi^3_\su$, and every operator in the super-WZW model can be written as a product of a parafermion operator and an operator from the super-$U(1)$ theory.  
In particular, the $\sutwo$ primary field $\Phihat_{\jsu \msu \bmsu}^{\su}$ can be decomposed as
\be
\label{supf app}
\Phihat^\su_{\jsu \msu \bmsu} = \Psihat^\su_{\jsu \msu \bmsu}
\,\exp\Bigl[i\frac2{\sqrt\nfive}\Bigl(\msu \cY_{\!\su}+\bmsu \bar\cY_{\!\su}\Bigr)\Bigr] ~.
\ee
The conformal dimension of the $\sutwo$ primary $\Phihat_{\jsu \msu \bmsu}^{\su}$ decomposes as
\be
\label{supfspec-app}
h(\Psihat^\su_{\jsu \msu \bmsu}) = \frac{\jsu (\jsu+1)-\msu^2}{\nfive} 
~~,~~~~
\bar h(\Psihat^\su_{\jsu \msu \bmsu}) = \frac{\jsu (\jsu+1)-\bmsu^2}{\nfive} 
\ee
with the rest made up by the dimension of the $\cY,\bar\cY$ exponentials.  

The shift $\msu \to (\msu \!+\!\frac12\nfive \wsu)$ in the $\cY$ exponential in $\Phihat_{\jsu \msu \bmsu}^{\su}$ in~\eqref{supf app} defines the left spectral flow of interest to us here.  The states flowed in this way have a shifted exponential but the same underlying superparafermion state; their conformal dimensions are
\be
h\bigl(\Psihat^{(\wsu,\bwsu)}_{\jsu \msu \bmsu}\bigr) = \frac{\jsu
  (\jsu+1)}{\nfive} + \msu \wsu +\frac{\nfive}{4} \wsu^2
\ee
and similarly for the right-handed spectral flow.


\subsection{\texorpdfstring{$\sltwo$}{}}
\label{sec:sltwocft}

The supersymmetric $\sltwo$ level $\nfive$ current algebra consists of currents $J_\sl^a$ and their fermionic superpartners $\psi_\sl^a$ having the OPE structure
\begin{align}
J_\sl^a(z)\,J_\sl^b(0) &\sim \frac{\frac12 \nfive \,h^{ab}}{z^2} + i\epsilon^{abc} \frac{J^\sl_c(0)}{z}
\nn\\
J_\sl^a(z)\, \psi_\sl^b(0) &\sim i\epsilon^{abc}\frac{\psi_c^\sl(0)}{z}
\\
\psi_\sl^a(z)\, \psi_\sl^b(0) &\sim \frac{h^{ab}}{z}
\nn
\end{align}
with the Killing metric $h^{ab} = {\rm diag}(+1,+1,-1)$.  One can similarly define a set of ``bosonic'' $\sltwo$ level $\nfivehat\!=\!\nfive\!+\!2$ currents $j_\sl^a$ that commute with the fermions,
\be
j_\sl^a = J_\sl^a + \frac i2\epsilon^{abc}\psi^\sl_b \psi^\sl_c  ~.
\ee
The primary fields $\Phihat_{\jsl \msl \bmsl}^{\sl}$ of the current algebra have conformal dimensions
\be
h = \bar h = -\frac{\jsl (\jsl-1)}{\nfive} ~.
\ee

As before the primary operators of the supersymmetric theory are built by combining primaries $\Phi^\sl_{\jsl \msl \bmsl}$ of the level $\nfivehat$ bosonic current algebra with level minus two primaries of the fermions $\One,\Sigma,\psi$.  The highest weight fields are similarly built by tensoring highest weight fields in the bosonic theory with one of these three, with the remaining operators of the zero mode multiplet obtained through the action of the zero modes of the total current $J^-_\sl$.  In building massless string states, one again uses the purely bosonic highest weight operator for NS sector states whose polarization does not lie along $\sltwo$, and the highest weight operator with a fermion attached when the polarization does lie along $\sltwo$; and Ramond sector operators involve $\Sigma$, with the various spinor polarizations reached through the action of the zero modes of $\psi^-_\sl,\bar\psi^-_\sl$. Again, in the type II string, the choice of fermion decoration is independent on left and right, and there is a chiral GSO projection onto odd total fermion number in the matter sector.

These operators also have a superparafermion decomposition under the current $J_\sl^3$%
~\cite{Dixon:1989cg,Griffin:1990fg,Dijkgraaf:1991ba}%
\footnote{Again our notation here largely follows~\cite{Martinec:2001cf}, see also~\cite{Giveon:2015raa}.}
obtained by extracting the dependence on $J^3_\sl, \bar J^3_\sl$.  To this end, one bosonizes the currents
\begin{align}
j^3_\sl=i\sqrt{\nfivehat}\,\partial Y_\sl 
~&,~~~~
\psi^+_\sl \psi^-_\sl = i\sqrt2\,\partial H_\sl ~,
\nn\\
J^3_\sl=i\sqrt{\nfive}\,\partial \cY _\sl
~&,~~~~
J_\cR^\sl = \frac\nfivehat\nfive \psi^+_\sl\psi^-_\sl + \frac2\nfive j_3^\sl
= i\sqrt{\frac{2\nfivehat}{\nfive}} \, \partial \cH_\sl ~,
\end{align}
and similarly for the right-movers.  Note that the boson $\cY$, $\bar\cY$ is timelike.  The $\sltwo$ primary field $\Phihat_{jm\mbar}$ can then be decomposed as
\begin{equation}
\label{slpf app}
\Phihat^\sl_{\jsl \msl \bmsl} = \Psihat^\sl_{\jsl \msl \bmsl }
\,\exp\Bigl[i\frac2{\sqrt\nfive}\Bigl(\msl \cY_{\!\sl}+\bmsl\bar\cY_{\!\sl}\Bigr)\Bigr] ~.
\end{equation}
The conformal dimension of the $\sltwo$ primary $\Phihat_{\jsl \msl \bmsl}^{\sl}$ decomposes as
\begin{equation}
\label{slpfspec-app}
h(\Psihat^\sl_{\jsl \msl \bmsl}) = \frac{-\jsl (\jsl-1)+\msl^2}{\nfive} 
~,~~~~
\bar h(\Psihat^\sl_{\jsl \msl \bmsl}) = \frac{-\jsl (\jsl-1)+\bmsl^2}{\nfive} ~,
\end{equation}
with the rest made up by the dimension of the $\cY,\bar\cY$
exponentials.  Again the fields $\Psihat^\sl_{\jsl \msl \bmsl}$ commute with the current $J^3_\sl$, and so are the natural building blocks for representations of the gauged theory.
The shift of the $J^3_\sl$ charge $\msl \to (\msl \!+\!\frac12 \nfive \wsl)$ leads to the flowed conformal dimension 
\be
h\bigl(\Psihat^{(\wsl,\bwsl)}_{\jsl \msl \bmsl}\bigr) = -\frac{\jsl
  (\jsl-1)}{\nfive} - \msl \wsl - \frac{\nfive}{4}  \wsl^2 \;.
\ee

Unitary representations of bosonic $\sltwo$ current algebra are as follows. One has the principal discrete series (on both left and right)
\begin{equation}
\cD_j^+ = \bigl\{ \ket{j,m}~\bigl| ~  j\in\IR_+\, ;~~ m\!=\! j+n\, ,~~n\in\IN \bigr\}
\end{equation}
and its conjugate 
\begin{equation}
\cD_j^- = \bigl\{ \ket{j,m}~\bigl| ~  j\in\IR_+\, ;~~ m\!=\! -(j+\bar n)\,
,~~\bar n\in\IN \bigr\} \,,
\end{equation}
restricted to the range
\be
\frac 12 \le j < \frac{\nfive+1}{2} ~;
\end{equation}
in addition one has the continuous series representations $\cC_j^\alpha$ (again on both left and right)
\begin{equation} \label{eq:cont-series-app}
\cC_j^\alpha = \bigl\{ \ket{j,m}~\bigl| ~  j\!=\!
\coeff12(1+i\nu)\, ,~~\nu \in\IR\, ;~~m\!=\! \alpha+n\, ,~~n\in\IZ\, ,~~0\!\le\alpha\!<1\in\IR \bigr\} ~.
\end{equation}

An important property of the representations results from the duality between the supersymmetric $\frac{\sltwo}{\uone}$ coset sigma model and $\cN\!=\!2$ Liouville theory (see~\cite{Giveon:2016dxe} and references therein).  A general version of this duality posits an isomorphism between the discrete series affine representations of the underlying bosonic WZW model
\be
\label{fieldident}
\bigl(\cD^-_j\bigr)^{w=\bar w=0} \equiv \bigl(\cD^+_{ \jdual }\bigr)^{w=\bar w= -1} ~,~~~~~~ \jdual \equiv \frac \nfive2-j +1~,
\ee 
which can be extended to the supersymmetric theory.
The quantum numbers of states in these two representations are equal, and embody the nature of the duality.  In particular, the isomorphism relates the two conformal operators
\be
(J_{-1}^+) (\bar J_{-1}^+)  \Phi_{1, -1, -1}^{\sst(0,0)}
~~~\longleftrightarrow~~~
\Phi_{\frac \nfive2,\frac\nfive2+1,\frac\nfive2+1}^{\sst(-1,-1)}
\ee
in the bosonic $\sltwo$ theory (where we have dropped the `$\sl$' decoration to reduce clutter).  One thus has an equivalence between the vertex operator that has the leading large $\rho$ asymptotics of the metric on the $\frac\sltwo\uone$ coset, and a winding tachyon condensate.  In the supersymmetric theory, the corresponding dual operators are the background metric of the supersymmetric coset, and the superpotential of $\cN=2$ Liouville theory.  This equivalence means that if one of these operators is condensed in the background of the model, then so is the other, since they are dual versions of the same object.  The winding tachyon condensate dominates the near-source dynamics, as discussed in Section~\ref{sec:Correlators}, following~\cite{Giveon:2015cma,Giveon:2016dxe}.

This field identification comes into play in the analysis of the winding spectrum of Section~\ref{sec:boundandwound}.
Note that in Eq.\;\eqref{BoundStateQnums}, if one chooses the other branch of the square root, then for $\cD^-$ representations the expression for $\Jsl$ has the form
\be
\Jsl  \sim \frac{1}{k} \Bigl[ w_y + \sqrt{ 2k^2\Ntot +2k\ntot w_y + {\X}\,}\, \Bigr]
\ee
where the details of $\X$ are not so important.  The point is that the energy 
\be
E \sim \frac{\nfive}{k^2\Ry}\Bigl[
\bigl(s(s\tight+1)+\bar s(\bar s\tight+1)\bigr) w_y - k \ntot - \sqrt{ 2k^2\Ntot +2k\ntot w_y + {\X}\,}\, \Bigr]
\ee
is below threshold relative to the background, and becoming more and more negative as we excite oscillators on the string.  However, we can apply the identification~\eqref{fieldident} which maps the state to a $\cD^+$ representation, with $\sltwo$ winding $\wsl=-1$.  Then using gauge spectral flow to return to our gauge choice $\wsl=0$, the winding is shifted to $w_y-k$, and one finds that the state has the branch of the square root chosen in~\eqref{BoundStateQnums}.
The energy is still below threshold, but now relative to that of an {\it antiwound} string, and growing more negative with increasing oscillator number; following the discussion at the beginning of Section~\ref{sec:winding}, the vertex operator with these quantum numbers is an {\it annihilation operator} for a string with the opposite sign of energy and conserved charges.  This is why in Section~\ref{sec:boundandwound} we only considered the other branch of the square root.


\newpage
\section{Conventions}
\label{app:conventions}

The conventions in this paper are chosen to agree with those
of \cite{Jejjala:2005yu,Chakrabarty:2015foa}. In this appendix we
record the relation to the conventions used in \cite{Martinec:2017ztd} (MM)
and in \cite{Giusto:2012yz} (GLMT). 

If we set $\m = \n + 1$, the JMaRT solution reduces to the
supersymmetric solution of
\cite{Giusto:2004id,Giusto:2004ip,Jejjala:2005yu,Giusto:2012yz}. In GLMT~\cite{Giusto:2012yz}, the angular
momenta
\begin{equation}\label{J_GLMT}
J_{\phi}^{GLMT} = \gamma_2 {N}\, ,\quad J_{\psi}^{GLMT} = \gamma_1 {N}
\end{equation} 
are given in terms of two parameters $\gamma_1$, $\gamma_2$,
related to the left spectral flow quantum number $s$ and the orbifold
order $k$ as
\begin{equation}
\gamma_1 = - \frac{s}{k} \, ,\quad \gamma_2 = \frac{s+1}{k} \, .
\end{equation}
The left/right quantities are defined by 
\begin{equation} \label{eq:j-glmt}
 J_L^{GLMT} = \frac12
(J_{\phi}^{GLMT}-J_{\psi}^{GLMT} )\, ,\quad   J_R^{GLMT} = \frac12
(J_{\phi}^{GLMT}+J_{\psi}^{GLMT} ) \, . 
\end{equation}
Setting $\m=s+1$, $\n = s$ in our solution, we find (after the standard conversion of units, see e.g.~\cite{Chakrabarty:2015foa,Bena:2016ypk,Bena:2017xbt})
\begin{equation}
J_{\phi} =\n \frac{N}{k} = s\frac{N}{k} \, ,\quad
J_{\psi}= -\m\frac{N }{k} = -(s+1)\frac{N}{k} \,
,
\end{equation} 
which differs from \eqref{J_GLMT}. To map to the GLMT expressions, one can relate
\be \label{eq:phi-psi-map-glmt}
\phi \;=\; -\psi_{GLMT} \,, \qquad \psi \;=\; -\phi_{GLMT} 
\ee
which effectively
exchanges $\m$ and $\n$. In the null constraints \eqref{jmartconstr}, this
corresponds to changing the sign of $\bar J^3_{\su}$ (and thus $\bar{m}_{\su}$) or $r_2$. 
 Note that we define the left/right angular momenta as
\begin{equation}
J^3_{\su} = J_L = \frac12 (J_{\phi} - J_{\psi})  = (\m+\n)\frac{N}{2k}\, ,\quad \bar{J}^3_{\su} = J_R = -\frac12 (J_{\phi}
+ J_{\psi})  = (\m - \n) \frac{N}{2k}\, ,
\end{equation}
which is consistent with \eq{eq:j-glmt} combined with \eq{eq:phi-psi-map-glmt}, such that both the present work and GLMT match the spacetime CFT quantities \eqref{JLandJR} with
$J_L = m_L$, $J_R = m_R$. 

In addition, compared to MM we have changed the sign of $\phi$ and
$\sigma$ in \eqref{groupelements} in order for the sigma model metric
to match with the JMaRT solution \eqref{eq:CFTmetric}.




\newpage

\bibliographystyle{JHEP}      

\bibliography{microstates}



\end{document}